\documentclass[preprint,pre]{revtex4}
\usepackage{graphicx}
\usepackage{lipsum}
\usepackage{float}
\usepackage{amsmath}
\usepackage{amssymb}
\usepackage{mathtools}
\include{mathmacros}

\begin{document}

\title{
Analysis of the convergence of the degree distribution of contracting 
random networks towards a Poisson distribution using the relative entropy
}

\author{Ido Tishby, Ofer Biham and Eytan Katzav}

\affiliation{
Racah Institute of Physics, 
The Hebrew University, 
Jerusalem 9190401, Israel
}

\begin{abstract}

We present analytical results for the structural evolution of 
random networks undergoing contraction processes via 
generic node deletion scenarios, namely,
random deletion, preferential deletion
and propagating deletion.
Focusing on configuration model networks,
which exhibit a given degree distribution $P_0(k)$
and no correlations,
we show using a rigorous argument
that upon contraction the degree distributions of these
networks converge towards a
Poisson distribution.
To this end, we
use the relative entropy $S_t=S[P_t(k) || \pi(k|\langle K \rangle_t)]$ 
of the degree distribution $P_t(k)$ of the contracting
network at time $t$ with respect to the corresponding Poisson distribution
$\pi(k|\langle K \rangle_t)$ with the same mean degree $\langle K \rangle_t$
as a distance measure between $P_t(k)$ and Poisson.
The relative entropy 
is suitable as a distance measure since it
satisfies $S_t \ge 0$ for any degree
distribution $P_t(k)$, while equality is obtained only for 
$P_t(k) = \pi(k|\langle K \rangle_t)$.
We derive an equation for the time derivative $dS_t/dt$
during network contraction 
and show that the relative entropy 
decreases monotonically to zero during the contraction process. 
We thus conclude that the degree distributions of
contracting configuration model networks 
converge towards a Poisson distribution.
Since the contracting networks remain uncorrelated,
this means that their structures converge towards an
Erd{\H o}s-R\'enyi (ER) graph structure,
substantiating earlier results obtained using
direct integration of the master equation and computer simulations
[I. Tishby, O. Biham and E. Katzav, 
{\it Phys. Rev. E} {\bf 100}, 032314 (2019)].
We demonstrate the convergence for configuration model networks with 
degenerate degree distributions (random regular graphs), exponential
degree distributions and power-law degree distributions (scale-free networks).

\end{abstract}

\maketitle

\section{Introduction}

Complex network architectures and dynamical processes
taking place on them play a central role in current research
\cite{Havlin2010,Newman2010,Estrada2011}.
Since the 1960s,
mathematical studies of networks
were focused on model systems  
such as the 
Erd{\H o}s-R\'{e}nyi (ER) network
\cite{Erdos1959,Erdos1960,Erdos1961},
which exhibits a Poisson degree distribution 
of the form
$\pi(k|c) = e^{-c} c^k /k!$,
where $c$ is the mean degree
\cite{Bollobas2001}.
In an ER network of $N$ nodes, 
each pair of nodes is connected with probability $p$,
where $p=c/(N-1)$.
In fact, ER networks 
form a maximum entropy ensemble under the constraint
that the mean degree is fixed
\cite{Bauer2002,Bogacz2006,Bianconi2008,Bianconi2009}.
In the 1990s, the growing availability of data on large biological, social and
technological networks revolutionized the field.
Motivated by the observation that the World Wide Web 
\cite{Albert1999} and 
scientific citation networks 
\cite{Redner1998}
exhibit power-law degree distributions,
Barab\'asi and Albert (BA) introduced a simple model
that captures the essential growth dynamics of such networks
\cite{Barabasi1999,Albert2002}.
A key feature of the BA model is the
preferential attachment mechanism, namely, the tendency of new nodes to
attach preferentially to high degree nodes.
Using mean-field equations and computer simulations
it was shown that the combination of growth and preferential attachment leads to the
emergence of scale-free networks with power-law degree distributions
\cite{Barabasi1999}. 
This result was later confirmed and generalized using a more rigorous
formulation based on the master equation
\cite{Krapivsky2000,Dorogovtsev2000}.
It was subsequently found that a large variety of empirical
networks exhibit such scale-free structures,
which are remarkably different from ER networks 
\cite{Albert2002,Barabasi2009}. 

In many of these networks the growth phase is not likely to proceed indefinitely.
Moreover, networks may be exposed to node deletion processes due to
node failures, attacks and epidemics, which may eventually halt the expansion
phase and induce the contraction and eventual collapse of the network.
Since network growth is a kinetic nonequilibrium processes,
it is not a reversible process, namely, the contraction process 
is not the same as the growth process when played backwards in time.
A particularly interesting example of the contraction phase can be seen in
the field of social networks. Such networks may lose users due to loss of
interest, concerns about privacy or due to their migration to other social networks
\cite{Torok2017,Lorincz2019}.
Another example of great practical importance is the cascading failure
of power-grids
\cite{Daqing2014,Schafer2018}.
Infectious processes such as epidemics that spread in a network 
\cite{Satorras2001,Satorras2015}
lead to the
contraction of the subnetwork of uninfected nodes and may thus be considered
as network contraction processes.
Similarly, network immunization schemes 
\cite{Satorras2002}
also belong to the class of network contraction
processes because they induce the contraction of the subnetwork of 
susceptible nodes.

Three generic scenarios of network contraction were identified: 
the scenario of random node deletion that describes the
random, inadvertent failure of nodes, the 
scenario of preferential node deletion that describes intentional attacks that
are more likely to focus on highly connected nodes
and the scenario of propagating node deletion that describes viral and
infectious processes that spread like epidemics. 
It was found that scale-free networks are resilient to attacks targeting
random nodes, but are vulnerable to attacks that target high degree nodes or hubs. 
Using the framework of percolation theory, 
it was shown that when the number of deleted nodes 
exceeds some threshold, the network breaks down into disconnected components
\cite{Albert2000,Cohen2000,Cohen2001,Braunstein2016,Zdeborova2016}.
However, the evolution of the network structure 
throughout the contraction phase was not addressed.

In a recent paper we analyzed the 
structural evolution of networks during the contraction process
\cite{Tishby2019}.
To this end we derived a master equation for
the time dependence of the degree distribution
during network contraction via the random deletion, preferential 
deletion and the propagating deletion scenarios.
Using the relative entropy and the degree-degree correlation function
we showed that the ER graph structure,
which exhibits a Poisson degree distribution, is an asymptotic structure for
these network collapse scenarios, in analogy to the way in which the 
scale-free structure is an asymptotic solution for the
preferential attachment growth scenario.

In this paper we use the relative entropy to provide a rigorous proof
that the ER structure is an attractive solution for the three contraction 
scenarios. This means that the ER structure is a universal asymptotic 
structure for contracting networks.
For simplicity, we consider initial networks drawn from configuration model network ensembles
that exhibit a desired degree distribution $P_0(k)$ and no degree-degree correlations.
During the contraction process the degree distribution of the network evolves.
We denote the degree distribution at time $t$ by $P_t(k)$ and its mean
degree by $\langle K \rangle_t$.
We use the relative entropy 
$S_t=S[P_t(k) || \pi(k|\langle K \rangle_t)]$ 
as a distance measure between 
the degree distribution $P_t(k)$ of the contracting
network and the corresponding Poisson distribution
$\pi(k|\langle K \rangle_t)$ with the same mean degree $\langle K \rangle_t$.
Using this measure we obtain rigorous results for
the convergence of the degree distribution of contracting
networks towards a Poisson distribution.
To this end, we derive an equation for the time derivative $dS_t/dt$
of the relative entropy during network contraction.
This equation can be expressed in the 
form $dS_t/dt = \Delta_{\rm A}(t) + \Delta_{\rm B}(t)$.
We show that $\Delta_{\rm A}(t) < 0$ for any degree distribution.
We also show that $\Delta_{\rm B}(t) < 0$ 
for degree distributions whose tails decay more slowly 
than the tail of the Poisson distribution with the same mean degree.
This condition is generically satisfied by the heavy-tail distributions that emerge
from network growth processes.
In contrast, in networks that exhibit narrow degree distributions
the $\Delta_{\rm B}(t)$ term turns out to be small and has little 
effect on the convergence, which is dominated by $\Delta_{\rm A}(t)$.
This implies that the relative entropy 
decreases monotonically during the contraction process.
Since the relative entropy satisfies $S_t \ge 0$ for any degree
distribution $P_t(k)$, while equality is obtained only for 
$P_t(k) = \pi(k|\langle K \rangle_t)$ 
we conclude that the degree distributions of contracting networks 
converge towards a Poisson distribution.
This conclusion is corroborated by the fact that the relative
entropy provides an upper bound for the total variation distance,
which is a standard measure of the difference between 
probability distributions.
We demonstrate the convergence for configuration model networks with a
degenerate degree distribution (random regular graphs), exponential
degree distribution and power-law degree distribution (scale-free networks).

The paper is organized as follows.
In Sec. II we present the three generic network contraction scenarios 
studied in this paper.
In Sec. III we present the master equation 
and show that the Poisson distribution is a solution of
the master equation for the three contraction scenarios.
In Sec. IV we present the relative entropy and express it in
terms of the Shannon entropy and the cross-entropy. 
In Sec. V we present rigorous results showing that the
relative entropy decays to zero in any of the three contraction scenarios.
In Sec. VI we present analytical results and computer simulations for
the contraction of configuration model networks with a degenerate degree
distribution (random regular graphs), an exponential degree distribution 
and a power-law degree distribution (scale-free networks).
The results are discussed in Sec. VII and summarized in Sec. VIII.

\section{Network contraction processes}

We consider network contraction processes in which at each time
step a single node is deleted together with its links.
The initial network consists of $N_0$ nodes, so at time $t$ the
network size is reduced to $N_t = N_0 - t$ nodes.
The deletion of a node of degree $k$, whose neighbors are of
degrees $k'_i$, $i=1,2,\dots,k$, 
eliminates the deleted node from the degree sequence
and reduces the degrees of its neighbors to 
$k'_i-1$, $i=1,2,\dots,k$. 
The node deleted at each time step is selected randomly.
However, the probability of a node to be selected for deletion may depend on
its degree, according to the specific network contraction scenario.
Here we focus on three generic scenarios of network contraction:
the scenario of random node deletion that describes the
random, inadvertent failure of nodes, the 
scenario of preferential node deletion that describes intentional attacks that
are more likely to focus on highly connected nodes
and the scenario of propagating node deletion that describes cascading failures and
infectious processes that spread throughout the network. 

In the random deletion scenario, at each time step a random 
node is selected for deletion.
In this scenario each one of the nodes in the network 
at time $t$ has the same probability to be
selected for deletion, regardless of its degree.
Since at time $t$ there are $N_t$ nodes in the network,
the probability of each one of them to be selected for deletion is $1/N_t$.
In the preferential deletion scenario  
the probability of a node to be selected for deletion at  time $t$ 
is proportional to its degree at that specific time.
This means that the probability of a given node of degree $k$ to be 
deleted at time $t$ is $k/[N_t \langle K \rangle_t]$.
This is equivalent to selecting a random edge in the network and 
randomly choosing for deletion one of the two nodes at its ends.
In the propagating deletion scenario at each time
step the node to be deleted is randomly selected among the
neighbors of the node deleted in the previous time step.
In case that the node deleted in the previous time step does
not have any yet-undeleted neighbor we pick a random node,
randomly select one of its neighbors for deletion and continue
the process from there.

Here we focus on the contraction of undirected networks of initial size $N$, which are drawn from a
configuration model network ensemble 
with a given initial degree distribution $P_0(k)$ and no degree-degree correlations.
The degree distribution is bounded from above and below
such that $k_{\rm min} \le k \le k_{\rm max}$.
For example, the commonly used choice of $k_{\rm min}=1$
eliminates the possibility of isolated nodes in the network. 
Choosing $k_{\rm min}=2$ also eliminates the leaf nodes.
Controlling the upper bound is important in the case of  
fat-tail degree distributions such as power-law degree distributions.
The configuration model network ensemble is a maximum entropy ensemble
under the condition that the degree distribution $P(k)$ is imposed
\cite{Molloy1995,Molloy1998,Newman2001,Annibale2009,Roberts2011,Coolen2017}.
In such uncorrelated networks the deletion
of a node at time $t$ does not induce correlations between the remaining $N_t-1$ nodes.
Thus, upon deletion of a node from a configuration model 
network of size $N_t$, the resulting network remains a
configuration model network with a suitably adjusted degree distribution $P_{t+1}(k)$.

\section{The master equation and its Poisson solution}

Consider an ensemble of 
networks of size $N_0$
and degree distribution $P_0(k)$, with
mean degree $\langle K \rangle_0$. 
At each time step a single node is deleted from the network.
In addition to the primary effect of the loss of the deleted node,
the damage to the network also includes a secondary effect as 
each neighbor of the deleted node loses one link.
An intrinsic property of the secondary effect is that it is
always of a preferential nature. 
This is due to the fact that the probability of a node of degree $k'$
to be a neighbor of the deleted node is proportional to $k'$.
The number of nodes
in the network at time $t$ is 
$N_t = N_0 - t$.
The number of nodes of degree $k$ at time $t$ is denoted by $N_t(k)$,
where  
$\sum_{k} N_t(k) = N_t$.
The time dependent degree distribution is given by

\begin{equation}
P_t(k) = \frac{N_t(k)}{N_t}.
\label{eq:P_t}
\end{equation}

\noindent
The mean degree and the second moment of the degree distribution at time $t$ are denoted by
$\langle K^n \rangle_t$
where $n=1$ and $2$, respectively.

The master equation 
\cite{vanKampen2007,Gardiner2004}
for the temporal evolution of
the degree distribution $P_t(k)$
during network contraction processes was derived in Ref.
\cite{Tishby2019}.
To demonstrate the derivation of the master equation
we consider below the relatively simple case of random node deletion.
The time dependence of $N_t(k)$
depends on the primary effect, given by the probability that the node selected
for deletion is of degree $k$, as well as on the secondary effect  of node deletion
on neighboring nodes of degrees $k$ and $k+1$.
In random node deletion the probability that the node 
selected for deletion at time $t$ is of degree $k$ is
given by $N_t(k)/N_t$.
Thus, the rate at which $N_t(k)$ decreases due to the primary effect of 
the deletion of nodes of degree $k$  
is given by

\begin{equation}
R_t(k \rightarrow \varnothing) = \frac{N_t(k)}{N_t},
\label{eq:Rk}
\end{equation}

\noindent
where $\varnothing$ represents the empty set.
In case that the node deleted at time $t$ is of degree $k'$,
it affects $k'$ adjacent nodes, which lose one link each. 
The probability of each one of these $k'$ nodes
to be of degree $k$ is given by
$k N_t(k)/[ N_t \langle K\rangle_t ]$.
We denote by $W_t(k \rightarrow k-1)$ the expectation value of
the number of nodes of degree $k$ that lose a link at time $t$ and
are reduced to degree $k-1$.
Summing up over all possible values of $k'$,  
we find that the secondary effect of random node deletion on nodes of 
degree $k$ amounts to

\begin{equation}
W_t(k \rightarrow k-1) =  \frac{kN_t(k)}{N_t}.
\label{eq:Wk}
\end{equation}

\noindent
Similarly, the secondary effect on nodes of degree $k+1$ 
amounts to

\begin{equation}
W_t(k+1 \rightarrow k) =  \frac{ (k+1)N_t(k+1)}{N_t}.
\label{eq:Wk+1}
\end{equation}

\noindent
The time evolution of $N_t(k)$ can be expressed in terms
of the forward difference

\begin{equation}
\Delta_t N_t(k) = N_{t+1}(k) - N_t(k).
\end{equation}

\noindent
Combining the primary and the
secondary effects on the time dependence of $N_t(k)$ 
we obtain

\begin{equation}
\Delta_t N_t(k) =
- R_t(k \rightarrow \varnothing) + \left[ W_t(k+1 \rightarrow k) - W_t(k \rightarrow k-1) \right].
\label{eq:RWW}
\end{equation}

\noindent
Since nodes are discrete entities the process of node deletion
is intrinsically discrete. Therefore, the replacement of the forward difference
$\Delta_t N_t(k)$
by a time derivative of the form 
$d N_t(k)/dt$ involves an approximation.
The error associated with this approximation was evaluated in 
Ref. 
\cite{Tishby2019}.
It was shown that except for the limit of extremely narrow
degree distributions the error is of order $1/N_t^2$,
which quickly vanishes in the large network limit.
This means that the replacement of the forward difference by
a time derivative has little effect on the results,
and a clear technical advantage.

Inserting the expressions for $R_t(k \rightarrow \varnothing)$, $W_t(k \rightarrow k-1)$ and
$W_t(k+1 \rightarrow k)$ from Eqs.
(\ref{eq:Rk}), (\ref{eq:Wk}) and (\ref{eq:Wk+1}), respectively 
into Eq. (\ref{eq:RWW})
and replacing $\Delta_t N_t(k)$ by $d N_t(k)/dt$
we obtain

\begin{equation}
\frac{d}{dt} N_t(k) =
\frac{(k+1)[ N_t(k+1) - N_t(k) ]}{N_t}.
\label{eq:DeltaNtk}
\end{equation}

\noindent
The derivation of the master equation is completed by taking the
time derivative of Eq. (\ref{eq:P_t}), which is given by

\begin{equation}
\frac{d}{dt} P_t(k) = \frac{1}{N_t} \frac{d}{dt} N_t(k) - \frac{N_t(k)}{N_t^2} \frac{d}{dt} N_t.
\label{eq:dPt_Nt}
\end{equation}

\noindent
Inserting the time derivative of $N_t(k)$ from Eq. (\ref{eq:DeltaNtk})
into Eq. (\ref{eq:dPt_Nt})
and using the fact that
$d N_t/dt=-1$,
we obtain the master equation for the random deletion scenario,
which is given by

\begin{equation}
\frac{d}{dt} P_t(k)=
\frac{1}{N_t}
\left[ (k+1)P_t(k+1) - k P_t(k) \right].
\label{eq:dP(t)/dtRC0}
\end{equation}

\noindent
The derivation of the master equations for the preferential deletion 
and the propagating deletion scenarios can be performed along similar lines
\cite{Tishby2019}.
Interestingly, the resulting master equations for these three network contraction scenarios
can be written in a unified manner, in the form

\begin{equation}
\frac{d}{dt} P_t(k) 
= 
F_{\rm A}(t) + F_{\rm B}(t),
\label{eq:dP/dt}
\end{equation}

\noindent
where

\begin{equation}
F_{\rm A}(t)
= 
\frac{A_t}{N_t}
\left[ (k+1) P_t(k+1) - k P_t(k) \right]
\label{eq:dP/dtA}
\end{equation}

\noindent
accounts for the secondary effect on the neighbors of the deleted node,
which lose one link each,
while

\begin{equation}
F_{\rm B}(t)
= 
- 
\frac{B_t(k)}{N_t}
P_t(k)
\label{eq:dP/dtB}
\end{equation}

\noindent
accounts for the primary effect, namely, the loss of the deleted node
\cite{Tishby2019}.
The coefficients $A_t$ and $B_t(k)$ are given by

\begin{equation}
A_t = 
\left\{
\begin{array}{ll}
  1   &  {\rm \ \ \ \ \   random \ deletion} \\
  \frac{\langle K^2 \rangle_t }{\langle K \rangle_t^2}   
& {\rm \ \ \ \ \   preferential \ deletion}  \\
      \frac{\langle K^2 \rangle_t - 2 \langle K \rangle_t}{\langle K \rangle_t^2} & 
{\rm \ \ \ \ \  propagating \ deletion,}
\end{array}
\right.
\label{eq:A_t}
\end{equation}

\noindent
and

\begin{equation}
B_t(k) =
\left\{
\begin{array}{ll}
 0  &  {\rm \ \ \ \ \  random \ deletion}    \\
  \frac{k - \langle K \rangle_t}{\langle K \rangle_t}  
&  {\rm \ \ \ \ \    preferential \ deletion}    \\
  \frac{k - \langle K \rangle_t}{\langle K \rangle_t}  
& {\rm \ \ \ \ \   propagating \ deletion}.
\end{array}
\right.
\label{eq:B_t(k)}
\end{equation}

\noindent
The master equation consists of a set of coupled ordinary differential equations
for $P_t(k)$, $k=0,1,2,\dots,k_{\rm max}$,
or in other words it is a partial difference-differential equation.
In order to calculate the time evolution of the degree distribution $P_t(k)$ during
the contraction process one solves the master equation using direct numerical integration
\cite{Butcher2003},
starting from the initial network that consists of $N_0$ nodes whose degree distribution
is $P_0(k)$. For any finite network the degree distribution is bounded from above
by an upper bound denoted by $k_{\rm max}$, which satisfies
the condition $k_{\rm max} \le N_0-1$. Since the contraction process can only
delete edges from the remaining nodes and cannot increase the degree of any
node, the upper cutoff $k_{\rm max}$ is maintained throughout the contraction
process.

\begin{figure}
\begin{center}
\includegraphics[width=6.0cm]{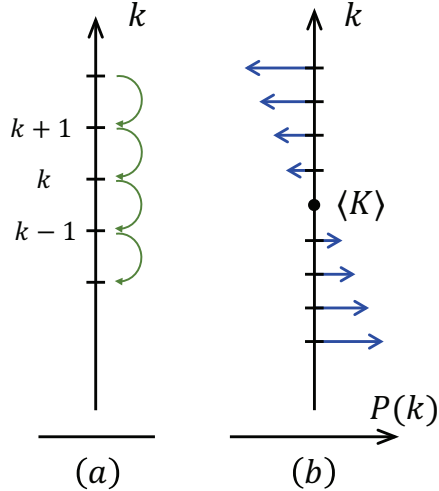}
\caption{
(Color online)
Illustration of the time dependence of the degree distribution 
$P_t(k)$ 
during network contraction processes,
described by the master equation (\ref{eq:dP/dt}).
(a) In the trickle-down term $F_{\rm A}(t)$, given by Eq. (\ref{eq:dP/dtA}),
the probability flows downwards step by step from degree $k+1$ to $k$
and from $k$ to $k-1$. This way high degree nodes become less probable
and low degree nodes become more probable as the contraction process evolves.
(b) In the redistribution term $F_{\rm B}(t)$, given by Eq. (\ref{eq:dP/dtB}),
for values of $k$ above the mean degree $\langle K \rangle_t$ the probability
$P_t(k)$ decreases at a rate proportional to $k-\langle K \rangle_t$,
while for values of $k$ below $\langle K \rangle_t$ the probability $P_t(k)$
increases at a rate proportional to $\langle K \rangle_t - k$.
Here the flow of probability is non-local in the $k$ axis, namely, probability
is lost at high degrees and instantaneously emerges at low degrees.
}
\label{fig:1}
\end{center}
\end{figure}

The $F_{\rm A}(t)$ term of the master equation, given by Eq. (\ref{eq:dP/dtA}),
is referred to as the trickle-down term
\cite{TrickleDown}.
This term represents
the step by step downwards flow of probability from high to low degrees.
This process is illustrated in Fig. 1(a).  
The coefficient $A_t$ of the trickle-down term depends on the network 
contraction scenario according to Eq. (\ref{eq:A_t}).
In the case of random node deletion $A_t=1$,
because the probability of a node to be selected for
deletion does not depend on its degree.
In the case of preferential node deletion $A_t$ is proportional to $\langle K^2 \rangle_t$
because the probability of a node to be deleted is proportional
to its degree $k$ while the magnitude of the secondary effect is also proportional to $k$.

The $F_{\rm B}(t)$ term of the master equation, given by Eq. (\ref{eq:dP/dtB}),
is referred to as the redistribution term.
As can be seen in Eq. (\ref{eq:B_t(k)}),
this term vanishes in the random deletion scenario.
However, in the preferential and propagating deletion scenarios
the redistribution term 
is negative for 
$k > \langle K \rangle_t$
and positive for
$k < \langle K \rangle_t$.
Thus the redistribution term decreases the probabilities
$P_t(k)$ for values of $k$ that are above the mean degree
and increases them for values of $k$ that are below the mean degree,
as illustrated in Fig. 1(b). 
The size of the redistribution term is proportional to the absolute value
$|k - \langle K \rangle_t|$,
which means that nodes of degrees that are much higher 
or much lower than $\langle K \rangle_t$ are most strongly affected
by this term.

Consider an ER network of $N_t$ nodes with mean degree 
$c_t$.
Its degree distribution follows
a Poisson distribution of the form

\begin{equation}
\pi(k|c_t) = \frac{ e^{-c_t} c_t^k }{k!}.
\label{eq:poisson}
\end{equation}

\noindent
The second moment of this degree distribution is equal to
$c_t(c_t + 1)$.
To examine the contraction process of ER networks
we start from an initial network of $N_0$ nodes
whose degree distribution follows a Poisson distribution $\pi(k|c_0)$,
where $c_0$ is the mean degree of the initial network.
Inserting 
$\pi(k|c_t)$
into the master equation (\ref{eq:dP/dt}) we find that 
the time derivative on the left hand side is given by

\begin{equation}
\frac{d}{dt} \pi(k|c_t) = 
- \frac{d c_t}{dt}
\left( 1 - \frac{k}{c_t} \right)  \pi(k|c_t),
\label{eq:dpi/dt1}
\end{equation}

\noindent
On the other hand,
inserting $\pi(k|c_t)$ on the right hand side of Eq. (\ref{eq:dP/dt}),
we obtain

\begin{equation}
\frac{d}{dt} \pi(k|c_t) =
\frac{A_t}{N_t} (c_t - k) \pi(k|c_t) 
- \frac{B_t(k)}{N_t} \pi(k|c_t),
\label{eq:dpi/dt2}
\end{equation}

%\noindent
%where
%upon replacement of the first and second moments
%$\langle K \rangle_t$ and $\langle K^2 \rangle_t$
%in Eqs. (\ref{eq:A_t}) and (\ref{eq:B_t(k)})
%by the corresponding values from an ER network,
%we obtain
%
%
%\begin{equation}
%A_t = 
%\left\{
%\begin{array}{ll}
%  1   &  {\rm \ \ \ \ \   random \ deletion} \\
%  \frac{c_t + 1}{c_t}   
%& {\rm \ \ \ \ \   preferential \ deletion}  \\
%     \frac{c_t - 1}{c_t} & 
%{\rm \ \ \ \ \  propagating \ deletion}
%\end{array}
%\right.
%\end{equation}
%
%\noindent
%and
%
%\begin{equation}
%B_t(k) =
%\left\{
%\begin{array}{ll}
% 0  &  {\rm \ \ \ \ \  random \ deletion}    \\
%  \frac{k - c_t}{c_t}  
%&  {\rm \ \ \ \ \    preferential \ deletion}    \\
%  \frac{k - c_t}{c_t}  
%& {\rm \ \ \ \ \   propagating \ deletion}.
%\end{array}
%\right.
%\end{equation}

\noindent
In order that $\pi(k|c_t)$ will be a solution of Eq. (\ref{eq:dP/dt}),
the right hand sides of Eqs. (\ref{eq:dpi/dt1}) and (\ref{eq:dpi/dt2}) 
must coincide.
In the case of random deletion this implies that

\begin{equation}
\frac{1}{c_t} \frac{d c_t}{dt} = - \frac{1}{N_t}.
\end{equation}

\noindent
Integrating both sides for $t'=0$ to $t$, we obtain
the solution 
$c_t = c_0 N_t/N_0$.
Repeating the analysis presented above for the
cases of preferential deletion and propagating deletion
it is found that $\pi(k|c_t)$
solves the master equation (\ref{eq:dP/dt})
for the three network contraction scenarios,
while
the mean 
degree, $c_t$ decreases linearly in time
according to

\begin{equation}
c_t = c_0 - R t,
\label{eq:c_linear}
\end{equation}

\noindent
where
the rate $R$
depends on the network contraction scenario, 
and is given by

\begin{equation}
R = 
\left\{
\begin{array}{ll}
 \frac{c_0}{N_0}   & {\rm \ \ \ \ \  random \ deletion}   \\
 \frac{c_0+2}{N_0}   &  {\rm \ \ \ \ \  preferential \ deletion}  \\
 \frac{c_0}{N_0}   & {\rm \ \ \ \ \  propagating \ deletion}.
\end{array}
\right.
\label{eq:c_t}
\end{equation}

\noindent
This means that an ER network exposed to
any one of the three contraction scenarios
remains an ER network at all times,
with a mean degree that decreases according to Eq.
(\ref{eq:c_linear}). 

\section{The relative entropy}

In order to establish that networks exposed to these contraction scenarios actually converge
towards the ER structure, it remains to show that the Poisson solution is attractive.
To quantify the convergence of $P_t(k)$,
whose mean degree is $\langle K \rangle_t$, 
towards a Poisson distribution,
we use the relative entropy
(also referred to as the Kullback-Leibler divergence),
defined by
\cite{Kullback1951}

\begin{equation}
S_t = S[P_t(k) || \pi(k|\langle K \rangle_t)] =
\sum_{k=0}^{\infty} P_t(k)
\ln \left[ \frac{P_t(k)}{\pi(k|\langle K \rangle_t)} \right],
\label{eq:S}
\end{equation}

\noindent
where 
$\pi(k|\langle K \rangle_t)$ 
is the Poisson distribution,
given by Eq. (\ref{eq:poisson}),
with the same mean degree
as $P_t(k)$, namely, $\langle K \rangle_t$.
The relative entropy $S_t$ is a distance measure between
the whole degree distribution $P_t(k)$ and the reference distribution $\pi(k|\langle K \rangle_t)$.
It also quantifies the added information associated with constraining the degree
distribution $P_t(k)$ rather than only the mean degree $\langle K \rangle_t$,
as nicely shown in Refs.
\cite{Annibale2009,Roberts2011,Coolen2017}.
The Poisson distribution is a proper reference distribution
for the relative entropy because it satisfies
$\pi(k|\langle K \rangle_t) > 0$ for all the non-negative integer values of $k$.
Using the log-sum inequality
\cite{Csiszar2004}, 
one can show that the
relative entropy is always non-negative and satisfies 
$S_t=0$ if and only if $P_t(k) = \pi(k|\langle K \rangle_t)$
\cite{Kullback1969,Cover2006}.
Therefore, $S_t$ can be used as
a measure of the distance between a given network and
the corresponding ER network with the same mean degree.

The relative entropy $S[P(k) || \pi(k|c)]$ of a degree distribution
$P(k)$ with mean degree $\langle K \rangle$ with respect to a 
Poisson distribution $\pi(k|c)$ with mean degree $c$
can be decomposed in the form

\begin{equation}
S[P(k) || \pi(k|c)] =
- S[P(k)] 
+ C[P(k) || \pi(k|c)]
\label{eq:SPpic}
\end{equation}

\noindent
where

\begin{equation}
S[P(k)] = - \sum_{k=0}^{\infty} P(k) \ln [ P(k) ]
\label{eq:Shannon}
\end{equation}

\noindent
is the Shannon entropy
\cite{Shannon1948}
of $P(k)$,
while 

\begin{equation}
C[P(k) || \pi(k|c)] =
- \sum_{k=0}^{\infty} P(k) \ln [ \pi(k|c) ], 
\label{eq:CPp}
\end{equation}

\noindent
is the cross-entropy 
\cite{Shore1980}
between $P(k)$ and $\pi(k|c)$.
The Poisson distribution $\pi(k|c)$ satisfies

\begin{equation}
\ln [ \pi(k|c) ]= -c + k \ln (c) - \ln (k!).
\label{eq:lnpi}
\end{equation}

\noindent
Inserting $\ln [\pi(k|c)]$ from Eq. (\ref{eq:lnpi}) into Eq. (\ref{eq:CPp}),
we obtain

\begin{equation}
S[P(k) || \pi(k|c)] = \sum_{k=0}^{\infty} P(k) \ln [ P(k) ]
+ c - \langle K \rangle \ln (c)
+ \sum_{k=0}^{\infty} \ln(k!) P(k).
\label{eq:Sk}
\end{equation}

\noindent
Eq. (\ref{eq:Sk}) provides the relative entropy of any degree distribution
$P(k)$ whose mean degree is $\langle K \rangle$, with respect to a Poisson distribution 
with mean degree $c$.
In order to find the value of $c$ for which 
$S[P(k) || \pi(k|c)]$ 
is minimal
we differentiate $S[P(k) || \pi(k|c)]$ with respect to $c$ and solve the equation

\begin{equation}
\frac{d}{dc} S[P(k) || \pi(k|c)] = 1 - \frac{\langle K \rangle}{c} = 0.
\end{equation}

\noindent
We find that $S[P(k) || \pi(k|c)]$ is minimized when the condition
$c = \langle K \rangle$
is satisfied.
This implies that for any degree distribution $P(k)$ with mean degree $\langle K \rangle$, 
the closest Poisson distribution $\pi(k|c)$,
in terms of the relative entropy,
is the Poisson distribution with mean degree 
$c=\langle K \rangle$.

Using the result discussed above, one can express the relative entropy
$S[P(k) || \pi(k|c)]$ in the form

\begin{equation}
S[P(k) || \pi(k|c)] = 
S[P(k) || \pi(k|\langle K \rangle)]
+
\delta S(c,\langle K \rangle)
\end{equation}

\noindent
where 
$S[P(k) || \pi(k|\langle K \rangle)]$
is the relative entropy of $P(k)$ with respect to a Poisson distribution
whose mean is $\langle K \rangle$, 
and

\begin{equation}
\delta S(c,\langle K \rangle) = \langle K \rangle \left[  \left( \frac{c}{\langle K \rangle} - 1 \right) 
-
\ln \left( \frac{c}{\langle K \rangle} \right) \right] 
\label{eq:DeltaS}
\end{equation}

\noindent
is the added entropy due to the difference between $c$ and $\langle K \rangle$.
Note that $\delta S(c,\langle K \rangle) \ge 0$ for any choice of $\langle K \rangle > 0$ and $c > 0$,
while $\delta S(c,\langle K \rangle) = 0$ only in the case that $c = \langle K \rangle$.

Going back to Eq. (\ref{eq:SPpic}), the
relative entropy 
$S[P(k) || \pi(k|\langle K \rangle)]$ 
can be expressed in the form

\begin{equation}
S[P(k) || \pi(k|\langle K \rangle)] = 
- S[P(k)] 
+ C[P(k) || \pi(k|\langle K \rangle)] ,
\end{equation}

\noindent
where $S[P(k)]$ is given by 
Eq. (\ref{eq:Shannon})
and

\begin{equation}
C[P(k) || \pi(k|\langle K \rangle)] =
\langle K \rangle - \langle K \rangle \ln (\langle K \rangle)
+ \sum_{k=0}^{\infty} \ln(k!) P(k).
\label{eq:Sk2}
\end{equation}

\noindent
To evaluate the last term in Eq. (\ref{eq:Sk2}) we recall that 
$\ln (0!) = \ln(1!) =0$,
while the $k=2$ term is $\ln (2) P(2)$.
For $k \ge 3$ we use the Stirling approximation
\cite{Olver2010}

\begin{equation}
\ln (k!) = \left( k + \frac{1}{2} \right) \ln (k) - k + \frac{1}{2} \ln (2 \pi).
\label{eq:Stirling}
\end{equation} 

\noindent
Inserting $\ln (k!)$ for $k \ge 3$ from Eq. (\ref{eq:Stirling}) into Eq. (\ref{eq:Sk2}) and
rearranging terms, we obtain

\begin{eqnarray}
C[P(k) || \pi(k|c)] &=& 
- \langle K \rangle \ln (\langle K \rangle) 
+ \sum_{k=2}^{\infty} \left( k + \frac{1}{2} \right) \ln (k) P(k)
\nonumber \\
&+& \frac{1}{2} \ln (2 \pi)
-\frac{1}{2} \ln (2 \pi) P(0)
+ \left[ 1 - \frac{1}{2} \ln (2 \pi) \right] P(1)
\nonumber \\
&+& \left[ 2 - \frac{3}{2} \ln (2) - \frac{1}{2} \ln (2 \pi) \right] P(2),
\label{eq:Skk}
\end{eqnarray}

\noindent
where the terms involving $P(0)$, $P(1)$ and $P(2)$
result from the adjustment of the summation due to the fact that 
Eq. (\ref{eq:Stirling}) is used only for $k \ge 3$.
Note that in the case of distributions in which $k_{\rm min} \ge 1$, one 
assigns $P(k) = 0$ for $0 \le k \le  k_{\rm min}-1$.
Using Eq. (\ref{eq:Skk}), the relative entropy of the degree distribution $P_t(k)$ of a contracting network
with respect to the corresponding Poisson distribution $\pi_t(k|\langle K \rangle_t)$ with the same mean 
degree $\langle K \rangle_t$,
is given by

\begin{eqnarray}
S_t &=&
\sum_{k=0}^{\infty} P_t(k) \ln [ P_t(k) ]
- \langle K \rangle_t \ln (\langle K \rangle_t) 
+ \sum_{k=2}^{\infty} \left( k + \frac{1}{2} \right) \ln (k) P_t(k)
\nonumber \\
&+& \frac{1}{2} \ln (2 \pi)
-\frac{1}{2} \ln (2 \pi) P_t(0)
+ \left[ 1 - \frac{1}{2} \ln (2 \pi) \right] P_t(1)
\nonumber \\
&+& \left[ 2 - \frac{3}{2} \ln (2) - \frac{1}{2} \ln (2 \pi) \right] P_t(2).
\label{eq:Skk7}
\end{eqnarray}

\noindent
Eq. (\ref{eq:Skk7}) is used in order to evaluate the relative entropy
during the contraction process, where $P_t(k)$ is obtained either from
numerical integration of the master equation or from computer simulations.

\section{Convergence of the relative entropy}

In each of the network contraction scenarios,
the degree distribution $P_t(k)$ evolves in time according to 
the master equation
[Eq. (\ref{eq:dP/dt})].
As a result, the relative entropy $S_t$ of the network also evolves as the
network contracts.
The time derivative of $S_t$ is given by

\begin{equation}
\frac{d}{dt} S_t =
\sum_{k=0}^{\infty} 
\ln \left[ \frac{P_t (k)}{\pi(k|\langle K \rangle_t)} \right] 
\frac{d}{dt}P_t(k)
+
\sum_{k=0}^{\infty} 
\frac{d}{dt} P_t(k) 
-
\sum_{k=0}^{\infty}
\frac{P_t(k)}{\pi(k|\langle K \rangle_t)}
\frac{d}{dt} \pi(k|\langle K \rangle_t).
\label{eq:ds/dt_full}
\end{equation}

\noindent
Replacing the order of the summation and the derivative in
the second term on the right hand side of Eq.
(\ref{eq:ds/dt_full}),
we obtain

\begin{equation}
\sum_{k=0}^{\infty} 
\frac{d}{dt} P_t(k) =
\frac{d}{dt} \left[ \sum_{k=0}^{\infty} P_t(k) \right] =0.
\end{equation}

\noindent
Inserting the derivative $d \pi(k|\langle K \rangle_t)/dt$ 
from Eq. (\ref{eq:dpi/dt1})
into the third term on the right hand side of
Eq. (\ref{eq:ds/dt_full}),
we obtain

\begin{equation}
\sum_{k=0}^{\infty}
\frac{P_t(k)}{\pi(k|\langle K \rangle_t)}
\frac{d}{dt} \pi(k|\langle K \rangle_t) =
-\frac{d \langle K \rangle_t}{dt}
\sum_{k=0}^{\infty}
\left( 1-\frac{k}{\langle K \rangle_t} \right) P_t(k) = 0.
\end{equation}

\noindent
Since the second and third terms in
Eq. (\ref{eq:ds/dt_full}) vanish,
the time derivative of the relative entropy is simply given by

\begin{equation}
\frac{d}{dt} S_t
=
\sum_{k=0}^{\infty} 
\ln \left[ \frac{P_t(k)}{\pi(k|\langle K \rangle_t)} \right]
\frac{d}{dt} P_t(k).
\label{eq:ds/dt}
\end{equation}

\noindent
This is a general equation that applies to any network contraction scenario
in which the Poisson distribution $\pi(k|\langle K \rangle_t)$ is a solution.
The relative entropy satisfies $S_t \ge 0$ for any degree distribution $P_t(k)$.
It vanishes if and only if $P_t(k) = \pi(k|\langle K \rangle_t)$.
Therefore, in order to prove the convergence of the degree distribution $P_t(k)$
towards a Poisson distribution in a given network contraction scenario, 
one needs to show that for this scenario
$dS_t/dt < 0$.
To this end, we use Eq. (\ref{eq:ds/dt}), where 
we replace the derivative
$dP_t/dt$ by the right hand side of the master equation,
Eq. (\ref{eq:dP/dt}).

For the analysis below it is convenient to
express the time evolution of the relative
entropy, given by Eq. (\ref{eq:ds/dt}), 
in the form

\begin{equation}
\frac{d}{dt} S_t = \Delta_{\rm A}(t) + \Delta_{\rm B}(t),
\label{eq:dSAB}
\end{equation}

\noindent
where $\Delta_{\rm A}(t)$ emanates from the $F_{\rm A}(t)$ term
(trickle-down term) of the master equation and
$\Delta_{\rm B}(t)$ emanates from the $F_{\rm B}(t)$ term
(redistribution term).
The contribution of the trickle-down term 
to $dS_t/dt$ is given by

\begin{equation}
\Delta_{\rm A}(t) = 
\frac{A_t}{N_t}
\sum_{k=0}^{\infty} 
\ln \left[\frac{P_{t}\left(k\right)}{\pi(k|\langle K \rangle_t)}\right]
\left[ (k+1) P_t(k+1) - k P_t(k) \right],
\label{eq:DeltaA1}
\end{equation}

\noindent
where $A_t$ is given by Eq. (\ref{eq:A_t}),
and the contribution of the redistribution term is given by

\begin{equation}
\Delta_{\rm B}(t) = 
- \frac{B}{N_t}
\sum_{k=0}^{\infty} 
\ln \left[\frac{P_{t}\left(k\right)}{\pi_{t}\left(k| \langle K \rangle_t \right)}\right]
\left( \frac{k}{\langle K \rangle_t} - 1 \right)
P_t(k),
\label{eq:DB1}
\end{equation}

\noindent
where

\begin{equation}
B  =
\left\{
\begin{array}{ll}
 0  &  {\rm \ \ \ \ \  random \ deletion}    \\
  1  
&  {\rm \ \ \ \ \    preferential \ deletion}    \\
  1  
& {\rm \ \ \ \ \   propagating \ deletion}.
\end{array}
\right.
\end{equation}

\noindent
In order to show that the degree distribution of the contracting
network converges towards a Poisson distribution, 
one needs to show that during the
contraction process
$\Delta_{\rm A}(t) + \Delta_{\rm B}(t) < 0$.
Below we consider each one of these terms separately.
We show that in all the three network contraction scenarios
and for any initial degree distribution $P_0(k)$,
the trickle-down term satisfies
$\Delta_{\rm A}(t) < 0$ 
at all times during the contraction process.
For the redistribution term $\Delta_{\rm B}(t)$
we obtain a necessary and sufficient condition on
the instantaneous degree distribution $P_t(k)$
under which 
$\Delta_{\rm B}(t) < 0$.
The condition essentially states that 
$\Delta_{\rm B}(t) < 0$
for any degree distribution whose tail decays 
more slowly than the tail of the Poisson distribution,
which decays super exponentially.
This condition is generically satisfied by empirical networks,
which are formed via growth processes.
The degree distributions of such networks typically exhibit fat tails,
which decay much more slowly than Poisson.

\subsection{Convergence due to the trickle-down term}

To gain more insight on the structure of the $\Delta_{\rm A}(t)$ term,
given by Eq. (\ref{eq:DeltaA1}), it is useful to express it in the form

\begin{equation}
\Delta_{\rm A}(t) = 
\frac{A_t}{N_t}
\left\{
\sum_{k=0}^{\infty} 
\ln \left[
\frac{P_t(k)}{\pi(k|\langle K \rangle_t)}
\right]
(k+1) P_t(k+1) 
-
\sum_{k=1}^{\infty} 
\ln \left[
\frac{P_t(k)}{\pi(k|\langle K \rangle_t)}
\right]
k P_t(k) 
\right\}.
\label{eq:DeltaA1b}
\end{equation}

\noindent
Taking a factor of $\langle K \rangle_t$ out of the curly parentheses
and multiplying the numerators and denominators in the arguments of
the logarithmic functions by $k/\langle K \rangle_t$ (for $k \ge 1$), we obtain

\begin{eqnarray}
\Delta_{\rm A}(t) &=& 
\frac{A_t}{N_t}
\left\{\langle K \rangle_t + \ln[P_t(0)] \right\} P_t(1) 
\\
&+&
\frac{A_t \langle K \rangle_t}{N_t}
\left\{
\sum_{k=1}^{\infty} 
\ln \left[
\frac{\widetilde P_t(k)}{\pi(k-1|\langle K \rangle_t)}
\right]
\widetilde P_t(k+1) 
-
\sum_{k=1}^{\infty} 
\ln \left[
\frac{\widetilde P_t(k)}{\pi(k-1|\langle K \rangle_t)}
\right]
\widetilde P_t(k) 
\right\},
\nonumber
\label{eq:DeltaA1c}
\end{eqnarray}

\noindent
where

\begin{equation}
\widetilde P_t(k) = \frac{ k }{\langle K \rangle_t} P_t(k),
\label{eq:tP_t(k)}
\end{equation}

\noindent
is the degree distribution of nodes selected via a random edge in a random
network with degree distribution $P_t(k)$.
Similarly, the distribution

\begin{equation}
\pi(k-1|\langle K \rangle_t) = \frac{k}{\langle K \rangle_t} \pi(k|\langle K \rangle_t) 
\label{eq:pi(k-1)}
\end{equation}

\noindent
can be interpreted as the degree distribution of nodes selected via a random edge
in an ER network with a Poisson degree distribution of the form
$\pi(k|\langle K \rangle_t)$.

Rewriting $\widetilde P_t(k+1)$
in the form $[\widetilde P_t(k+1)/\widetilde P_t(k)] \widetilde P_t(k)$,
one can express the 
$\Delta_{\rm A}(t)$ term as a covariance of the form

\begin{eqnarray}
\Delta_{\rm A}(t) &=& 
\frac{A_t}{N_t}
\left\{   \langle K \rangle_t P_t(1) + \ln[P_t(0)]   P_t(1) 
- \frac{P_t(1)}{\langle K \rangle_t}
S[\widetilde P_t(k) || \pi(k-1|\langle K \rangle_t)]
\right.
\\
&+&
\left.
\widetilde {\mathbb E}_t \left[
  \frac{\widetilde P_t(k+1)}{\widetilde P_t(k)}  
\ln \left(
\frac{\widetilde P_t(k)}{\pi(k-1|\langle K \rangle_t)}
\right)
\right]
-
\widetilde {\mathbb E}_t \left[
\frac{\widetilde P_t(k+1)}{\widetilde P_t(k)}
\right]
\widetilde {\mathbb E}_t \left[
\ln \left(
\frac{\widetilde P_t(k)}{\pi(k-1|\langle K \rangle_t)}
\right)
\right]
\right\},
\nonumber
\label{eq:DeltaA1d}
\end{eqnarray}

\noindent
where 
$\widetilde {\mathbb E}_t[f(k)] = \sum_k f(k) \widetilde P_t(k)$.
In particular,

\begin{equation}
\widetilde {\mathbb E}_t \left[
\frac{\widetilde P_t(k+1)}{\widetilde P_t(k)}
\right]
=
\sum_{k=1}^{\infty}
\left( \frac{\widetilde P_t(k+1)}{\widetilde P_t(k)} \right)
\widetilde P_t(k) = 1 - \frac{P_t(1)}{\langle K \rangle_t}.
\end{equation}

\noindent
In order that the covariance will be negative,
in domains in which
$\widetilde P_t(k)$ 
is an increasing function [namely, $\widetilde P_t(k+1) > \widetilde P_t(k)$],
it should  be lower than the corresponding
Poisson distribution
[namely, $\widetilde P_t(k) < \pi(k-1|\langle K \rangle_t)$],
while in domains in which
$\widetilde P_t(k)$ 
is a decreasing function
it should be higher than the corresponding Poisson distribution.

In order to prove that $\Delta_{\rm A}(t) < 0$
for any degree distribution $P_t(k)$ at all stages of the contraction process
we rewrite Eq. (\ref{eq:DeltaA1}) in
the form

\begin{equation}
\Delta_{\rm A}(t) = \Delta_{\rm A}^{\rm P}(t) - \Delta_{\rm A}^{\pi}(t),
\label{eq:DeltaSplit}
\end{equation}

\noindent
where

\begin{equation}
\Delta_{\rm A}^{\rm P}(t) = 
\frac{A_t}{N_t}
\sum_{k=0}^{\infty} 
\ln \left[ P_{t}\left(k\right) \right]
\left[ (k+1) P_t(k+1) - k P_t(k) \right],
\label{eq:DAP}
\end{equation}

\noindent
and

\begin{equation}
\Delta_{\rm A}^{\pi}(t) = 
\frac{A_t}{N_t}
\sum_{k=0}^{\infty} 
\ln \left[\pi \left(k| \langle K \rangle_t \right)\right]
\left[ (k+1) P_t(k+1) - k P_t(k) \right].
\label{eq:DApi}
\end{equation}

\noindent
Separating the sum in Eq. (\ref{eq:DAP})
into two sums and replacing $k+1$ by $k$ in the first sum, we obtain

\begin{equation}
\Delta_{\rm A}^{\rm P}(t) = 
\frac{A_t}{N_t}
\left\{
\sum_{k=1}^{\infty} 
\ln \left[ P_{t}\left(k-1\right) \right]
k P_t(k)
- 
\sum_{k=1}^{\infty} 
\ln \left[ P_{t}\left(k\right) \right]
 k P_t(k)
\right\}.
\label{eq:DAPs}
\end{equation}

\noindent
Expressing the degree distribution $P_t(k)$ in terms of
$\widetilde P_t(k)$,

\begin{eqnarray}
\Delta_{\rm A}^{\rm P}(t) &=& 
\frac{A_t \langle K \rangle_t}{N_t}
\left\{
\sum_{k=1}^{\infty} 
\ln \left[ P_{t}\left(k-1\right) \right]
\widetilde P_t(k)
-\sum_{k=1}^{\infty} 
\ln \left[ \widetilde P_{t}\left(k\right) \right]
 \widetilde P_t(k) 
\right\} 
\nonumber \\
&+&
\frac{A_t}{N_t}
\sum_{k=1}^{\infty} \ln \left( \frac{k}{\langle K \rangle_t} \right) k P_t(k)
.
\label{eq:DAP2}
\end{eqnarray}

\noindent
Combining the first two terms 
in Eq. (\ref{eq:DAP2})
and splitting the last term, we obtain

\begin{equation}
\Delta_{\rm A}^{\rm P} (t) = 
- \frac{ A_t \langle K \rangle_t }{N_t} 
\sum_{k=1}^{\infty}
\widetilde P_t(k) \ln \left[ \frac{ \widetilde P_t(k) }{ P_t(k-1) } \right]
+ \frac{A_t}{N_t} 
\langle K \ln (K) \rangle_t
- \frac{A_t}{N_t} \langle K \rangle_t  \ln (\langle K \rangle_t).
\label{eq:DAP3}
\end{equation}

In order to evaluate $\Delta_{\rm A}^{\pi}$ we insert 

\begin{equation}
\ln [\pi(k| \langle K \rangle_t)] = - \langle K \rangle_t + k \ln (\langle K \rangle_t) - \ln (k!)
\end{equation}

into Eq. (\ref{eq:DApi})
and obtain

\begin{equation}
\Delta_{\rm A}^{\pi}(t) = 
\frac{A_t}{N_t}
\sum_{k=0}^{\infty} 
[- \langle K \rangle_t + k \ln (\langle K \rangle_t) - \ln(k!)]
\left[ (k+1) P_t(k+1) - k P_t(k) \right].
\end{equation}

\noindent
Carrying out the summation and
using the identity 

\begin{equation}
\ln (k!) = \ln [(k+1)!] - \ln (k+1),
\end{equation}

\noindent
we obtain

\begin{equation}
\Delta_{\rm A}^{\pi}(t) = 
\frac{A_t}{N_t}
\langle K \ln (K) \rangle_t
-
\frac{A_t}{N_t}
\langle K \rangle_t \ln (\langle K \rangle_t).
\label{eq:DApi2}
\end{equation}

\noindent
Inserting the results for $\Delta_{\rm A}^{\rm P}$ and
$\Delta_{\rm A}^{\pi}$, 
from Eqs. (\ref{eq:DAP3}) and (\ref{eq:DApi2}), respectively,
into Eq. (\ref{eq:DeltaSplit}),
we obtain

\begin{equation}
\Delta_{\rm A}(t) =
- \frac{A_t \langle K \rangle_t }{N_t}
S[ \widetilde P_t(k) || P_t(k-1) ]
\end{equation}

\noindent
where

\begin{equation}
S[ \widetilde P_t(k) || P_t(k-1) ] =
\sum_{k=1}^{\infty} 
\widetilde P_t(k) \ln \left[ \frac{ \widetilde P_t(k) }{ P_t(k-1) } \right]
\label{eq:Skkm1}
\end{equation}

\noindent
is the relative entropy of $\widetilde P_t(k)$ with respect to
$P_t(k-1)$. 
Note that Eq. (\ref{eq:Skkm1}) is valid only if
$P_t(k-1) > 0$ for all values of $k$ for which
$\widetilde P_t(k) > 0$. 
This means that the degree distribution should not have any gaps,
namely, values of $k'$ for which $P_t(k') =0$ while $P_t(k) > 0$
for any $k > k'$.
In practice, even if there are such gaps in the initial degree distribution
$P_0(k)$, they are quickly filled up due to the trickle-down term $F_{\rm A}(t)$ of the 
master equation, given by Eq. (\ref{eq:dP/dtA}).

Since the relative entropy must be positive, we find that
$\Delta_A(t) < 0$ for any degree distribution $P_t(k)$ that differs from $\pi(k|\langle K \rangle_t)$.
Actually, since the only distribution for which
$S[ \widetilde P_t(k) || P_t(k-1) ] = 0$ is the Poisson distribution,
this process can converge only to the Poisson distribution.
In the random deletion scenario, only the $\Delta_A(t)$ term contributes to the
time evolution of $S_t$, while the $\Delta_B(t)$ term vanishes.
This means that in the random deletion scenario
the distance between $P_t(k)$ and the corresponding Poisson
distribution $\pi(k|\langle K \rangle_t)$ with the same mean degree  
$\langle K \rangle_t$ decreases monotonically
at any stage during the contraction process.
In the preferential deletion and the propagating deletion scenarios
the convergence also depends on the $\Delta_{\rm B}(t)$ term,
which is considered below.

\subsection{Convergence due to the redistribution term}

In order to gain insight on the $\Delta_{\rm B}(t)$ term, 
we rewrite Eq. (\ref{eq:DB1})
in the form

\begin{equation}
\Delta_{\rm B}(t) = 
- \frac{B}{N_t}
\left\{
\sum_{k=1}^{\infty} 
\ln \left[\frac{P_{t}\left(k\right)}{\pi\left(k| \langle K \rangle_t \right)}\right]
\frac{k}{ \langle K \rangle_t }
P_t(k)
-
\sum_{k=0}^{\infty} 
\ln \left[\frac{P_{t}\left(k\right)}{\pi(k| \langle K \rangle_t )}\right]
P_t(k)
\right\}.
\label{eq:DB2}
\end{equation}

\noindent
Taking the factor of $1/\langle K \rangle_t$ out of the curly brackets, we obtain

\begin{equation}
\Delta_{\rm B}(t) = 
- \frac{B}{\langle K \rangle_t N_t}
\left\{
\sum_{k=1}^{\infty} 
k
\ln \left[\frac{P_{t}\left(k\right)}{\pi\left(k| \langle K \rangle_t \right)}\right]
P_t(k)
-
\langle K \rangle_t
\sum_{k=0}^{\infty} 
\ln \left[\frac{P_{t}\left(k\right)}{\pi(k| \langle K \rangle_t )}\right]
P_t(k)
\right\}.
\label{eq:DB3}
\end{equation}

\noindent
The expression in the curly brackets is, in fact, 
equal to the covariance 
between $k$ and $\ln[P_t(k)/\pi(k|\langle K \rangle_t)]$ under the
distribution $P_t(k)$, namely

\begin{equation}
\Delta_{\rm B}(t) = 
- \frac{B}{\langle K \rangle_t N_t}
\left\{
\left\langle
k
\ln \left[\frac{P_{t}\left(k\right)}{\pi(k| \langle K \rangle_t )}\right]
\right\rangle
-
\langle K \rangle_t
\left\langle
\ln \left[\frac{P_{t}\left(k\right)}{\pi(k| \langle K \rangle_t )}\right]
\right\rangle
\right\}.
\label{eq:DB4}
\end{equation}

\noindent
Therefore, in the case of distributions for which the correlation between
$k$ and $\ln[P_t(k)/\pi(k|\langle K \rangle_t)]$ is positive, the term in the
curly brackets is positive and $\Delta_{\rm B}(t) < 0$.
In this case the $\Delta_{\rm B}(t)$ term contributes to the convergence of
$P_t(k)$ towards a Poisson distribution. 
Such positive correlation essentially implies that for large values of $k$,
$P_t(k)$ tends to be larger than $\pi(k|\langle K \rangle_t)$, namely, 
it has a heavier tail than the Poisson distribution
with the same mean value.
Since network growth processes generically lead to fat tail distributions
such as the power-law distributions of scale-free networks, it is expected
that most empirical networks will exhibit a positive correlation between
$k$ and $\ln[P_t(k)/\pi(k|\langle K \rangle_t)]$.

In those cases in which the correlation between 
$k$ and $\ln[P_t(k)/\pi(k|\langle K \rangle_t)]$ is negative, the term in the
curly brackets is negative and $\Delta_{\rm B}(t)>0$.
In this case the $\Delta_{\rm B}(t)$ term works against the convergence
of $P_t(k)$ towards a Poisson distribution.
However, comparing the coefficients of $\Delta_{\rm A}(t)$ and $\Delta_{\rm B}(t)$
one finds that the coefficient of $\Delta_{\rm A}(t)$ is effectively larger by a factor
of $\langle K^2 \rangle/\langle K \rangle$ than the coefficient of $\Delta_{\rm B}(t)$.
Therefore, it is expected that the $\Delta_{\rm A}(t)$ term will be dominant and 
induce the convergence of $P_t(k)$ towards Poisson even in those cases
in which $\Delta_{\rm B}(t)>0$.

To gain more insight into the sign of $\Delta_{\rm B}(t)$
from a different perspective, we use
Eqs. (\ref{eq:tP_t(k)}) and (\ref{eq:pi(k-1)})
to express $\Delta_{\rm B}(t)$ 
of Eq. (\ref{eq:DB2})
in the form

\begin{equation}
\Delta_{\rm B}(t) = 
-
\frac{B}{N_t}
\sum_{k=1}^{\infty} 
\ln \left[\frac{\widetilde P_{t}\left(k\right)}{\pi\left(k-1| \langle K \rangle_t \right)}\right]
\widetilde P_t(k)
+
\frac{B}{N_t}
\sum_{k=0}^{\infty} 
\ln \left[\frac{P_{t}\left(k\right)}{\pi \left(k| \langle K \rangle_t \right)}\right]
P_t(k).
\label{eq:DeltaB}
\end{equation}

\noindent
The first sum in Eq. (\ref{eq:DeltaB}) is the
relative entropy of the degree distribution 
$\widetilde P_t(k)$ with respect to the shifted Poisson 
distribution $\pi(k-1| \langle K \rangle_t)$.
This is essentially a distance measure between the 
degree distribution of nodes selected preferentially in 
a network whose degree distribution is $P_t(k)$ and
the degree distribution of nodes selected preferentially
from the corresponding Poisson distribution with the
same mean degree.
The second term in Eq. (\ref{eq:DeltaB}) is the relative entropy of the 
degree distribution $P_t(k)$ with respect to the 
Poisson distribution $\pi(k|\langle K \rangle_t)$,
which is essentially a distance measure between $P_t(k)$ and $\pi(k|\langle K \rangle_t)$.
Thus, Eq. (\ref{eq:DeltaB}) can be written in the form

\begin{equation}
\Delta_{\rm B}(t) = 
- \frac{B}{N_t}
\left\{ S[\widetilde P_t(k) || \pi(k-1|\langle K \rangle_t)] - S[P_t(k) || \pi(k | \langle K \rangle_t)]
\right\}.
\label{eq:DeltaB3}
\end{equation}

\noindent
In the case that the degree distributions obtained for the preferential selection
are farther apart than the degree distributions obtained for random selection,
$\Delta_{\rm B}(t) < 0$, while in the opposite case $\Delta_{\rm B}(t) > 0$.

There is an important distinction between the two terms in Eq. (\ref{eq:DeltaB3}).
The second term is the relative entropy of $P_t(k)$ with respect to the
Poisson distribution $\pi(k| \langle K \rangle_t)$
with the same mean degree $\langle K \rangle_t$.
In contrast, the first term is the relative entropy of $\widetilde P_t(k)$ with
respect to the Poisson distribution $\pi(k-1|\langle K \rangle_t)$. 
The mean degree of
$\widetilde P_t(k)$ is

\begin{equation}
\langle \widetilde K \rangle_t = \frac{ \langle K^2 \rangle_t }{ \langle K \rangle_t },
\end{equation}

\noindent
while the mean degree of $\pi(k-1| \langle K \rangle_t)$ is $\langle K \rangle_t + 1$.
Therefore, Eq. (\ref{eq:DeltaB3}) can be written in the form

\begin{eqnarray}
\Delta_{\rm B}(t) 
&=& 
- \frac{B}{N_t}
\left\{ S[\widetilde P_t(k) || \pi(k-1|\langle \widetilde K \rangle_t - 1)] - S[P_t(k) || \pi(k | \langle K \rangle_t)]
\right\}
\nonumber \\
&-& \frac{B}{N_t}
\delta S(\langle \widetilde K \rangle_t,  \langle K \rangle_t + 1),
\label{eq:DeltaB4}
\end{eqnarray}

\noindent
where
$\delta S(\langle \widetilde K \rangle_t, \langle K \rangle_t+1)$
is given by Eq. (\ref{eq:DeltaS}).
This implies that $\Delta_{\rm B}(t) < 0$ as long as

\begin{equation}
S[\widetilde P_t(k) || \pi(k-1|\langle \widetilde K \rangle_t - 1)] > S[P_t(k) || \pi(k | \langle K \rangle_t)]
- \delta S(\langle \widetilde K \rangle_t, \langle K \rangle_t + 1).
\end{equation}

\noindent
Since 
$\delta S(\langle \widetilde K \rangle_t, \langle K \rangle_t+1)$ 
is always positive and its value increases as $P(k)$ becomes broader,
this condition is expected to be satisfied for any degree distribution 
that exhibits a heavy tail.
From our experience, degree distributions for which $\Delta_{\rm B} > 0$
are very special, usually hand-crafted for the mission.
In those cases, $\Delta_{\rm A}$, which is always negative, as proven above,
is much larger in absolute value than $\Delta_{\rm B}$.

\section{Contraction of networks with given initial degree distributions}

Here we apply the framework presented above 
to three examples of configuration model networks, 
with a degenerate degree distribution (also known as random regular graphs),
an exponential degree distribution and
a power-law degree distribution 
(scale-free networks).

\subsection{Random regular graphs}

A random regular graph (RRG) is a configuration model network in which all
the nodes are of the same degree, $k=c_0$, namely 

\begin{equation}
P_0(k) = \delta_{k,c_0}, 
\label{eq:deg}
\end{equation}

\noindent
where $c_0$ is an integer.
Here we consider the case of
$c_0 \ge 3$, in which the giant component encompasses the whole network.
In order to leave room for contraction into a non-trivial degree distribution,
we choose RRGs with $c_0 \gg 1$. Since in node deletion processes  
the degrees of nodes in the network are only reduced and never increase
it is clear that the range of
degrees of the contracted network will be limited to $0 \le k \le c_0$.
This means that in the case that the initial network is an RRG the tail of
the degree distribution of the contracted network will be truncated 
above $k=c_0$. Thus, the convergence towards Poisson is expected to
be relatively slow.

To evaluate the relative entropy of the initial RRG network with respect to
the corresponding Poisson distribution 
we insert the degenerate distribution of Eq. (\ref{eq:deg})
into Eq. (\ref{eq:S}).
We obtain the initial relative entropy

\begin{equation}
S_0 =  \ln \left[ \frac{1}{\pi(c_0|c_0)} \right].
\label{eq:Srrg}
\end{equation}

\noindent
Inserting the Poisson degree distribution 
into Eq. (\ref{eq:Srrg}) we obtain

\begin{equation}
S_0 = c_0 - c_0 \ln (c_0) + \ln (c_0!).
\end{equation}

\noindent
Using the Stirling approximation to evaluate $\ln (c_0!)$, 
we obtain

\begin{equation}
S_0 = 
\frac{1}{2} \ln (c_0) + \frac{1}{2}  \ln (2 \pi).
\label{eq:S0rrg}
\end{equation}

Below we analyze the convergence of a configuration model network
with a degenerate degree distribution towards an ER graph structure upon 
contraction. 
In particular, we calculate the time-dependent degree distribution
$P_t(k)$ during contraction and examine its convergence towards $\pi(k|\langle K \rangle_t)$.
To this end we perform direct numerical integration of the 
master equation (\ref{eq:dP/dt}) and computer simulations,
starting from a configuration model network
with a degree distribution
given by Eq. (\ref{eq:deg})
and evaluate the time-dependent relative entropy $S_t$.

\begin{figure}
\begin{center}
\includegraphics[width=6.0cm]{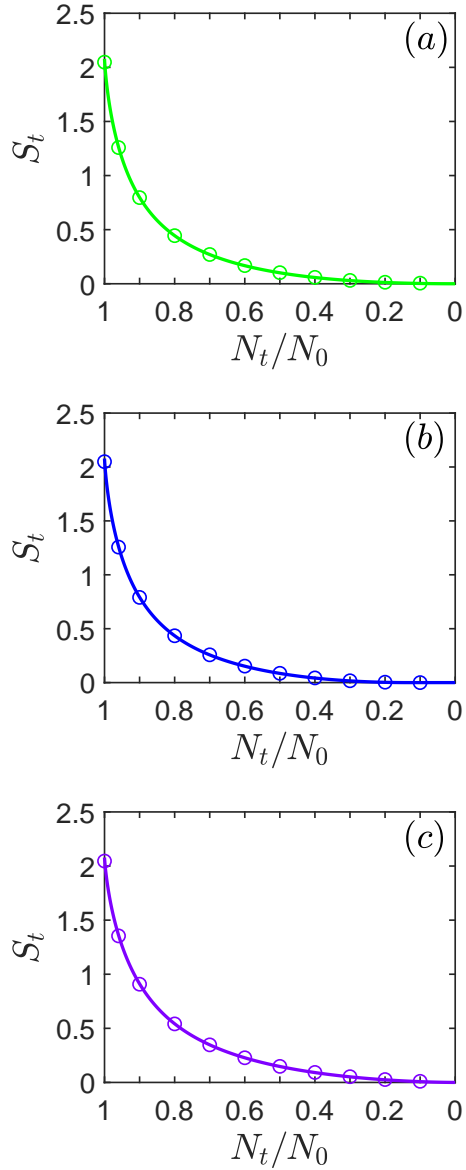}
\caption{
(Color online)
The relative entropy $S_t$ as a function of time 
for a random regular graph
of initial size $N_0=10^4$ and initial degree $c_0=10$
that contracts via random deletion (a), preferential deletion (b) and propagating
deletion (c), obtained from numerical integration of the master equation (solid lines).
In all three cases the relative entropy quickly decays,
which implies that the degree distribution of the contracting network
converges towards a Poisson distribution.
The master equation results 
are in excellent agreement with the results obtained from computer simulations (circles).
Also, the initial value $S_0 \simeq 2.08$ is in perfect agreement with 
the result obtained from Eq. (\ref{eq:S0rrg}).
}
\label{fig:2}
\end{center}
\end{figure}

In Fig. \ref{fig:2} we present the relative entropy $S_t$ as a function of time 
(represented by $N_t/N_0 = 1 - t/N_0$)
for a random regular graph of size $N_0=10^4$ with
a degenerate degree distribution in which all the nodes are of degree $c_0=10$,
that contracts via: (a) random node deletion; (b) preferential node deletion; and (c) propagating node deletion.
The results obtained from numerical integration of the master equation (solid lines)
are in excellent agreement with the results obtained from computer simulations,
namely, direct simulations of contracting networks (circles).
In all three cases the relative entropy quickly decays,
which implies that the degree distribution $P_t(k)$ of the contracting network
converges towards a Poisson distribution.
The decay rate of $S_t$ is comparable in all the three scenarios.
This implies that for extremely narrow degree distributions such as the degenerate
distribution the preferential and the propagating deletion scenarios do not exhibit
faster convergence than the random deletion scenario.

\begin{figure}
\begin{center}
\includegraphics[width=13.4cm]{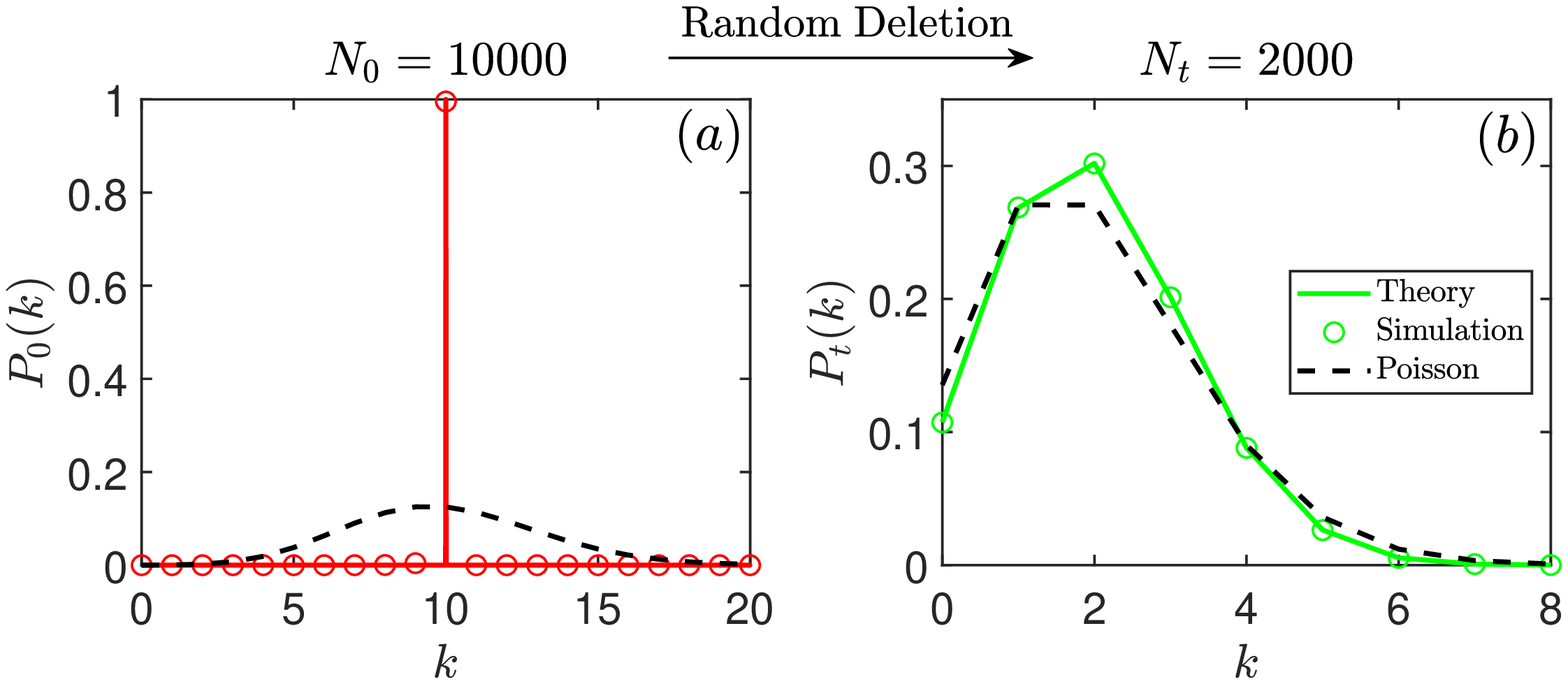}
\caption{
(Color online)
(a) The degree distribution $P_0(k)$ of a random regular graph (solid line)
in which all the nodes are of degree $c_0=10$.
The circles represent the degree sequence of a single network 
instance of $N_0=10^4$
nodes, which was used in the computer simulations.
The corresponding Poisson distribution
with the same mean degree is also shown (dashed line).
The network contracts via random node deletion.
(b) The degree distribution $P_t(k)$ of the contracted network
at time $t=8000$, when the network size is reduced to $N_t=2000$.
The results obtained from numerical integration of the master equation (solid line)
are in excellent agreement with the results obtained
from computer simulations (circles).
They are both very well converged towards the
corresponding Poisson distribution $\pi(k|\langle K \rangle_t)$
with the same mean degree $\langle K \rangle_t$ (dashed line).
}
\label{fig:3}
\end{center}
\end{figure}

In Fig. \ref{fig:3}(a) we present the degree distribution $P_0(k)$ of a random regular graph (solid line)
of size $N_0=10^4$ with a degenerate degree
distribution in which all the nodes are of degree $c_0=10$.
The corresponding Poisson distribution 
with the same mean degree 
$\langle K \rangle_0=c_0$ 
is also shown (dashed line).
Clearly, it is highly dissimilar to the degenerate distribution.
The random regular graph undergoes a network contraction process 
via the random node deletion scenario.
In Fig. \ref{fig:3}(b) we present the degree distribution $P_t(k)$ of the contracted network
at time $t=8000$, where the contracted network size is $N_t=2000$.
The results obtained from the numerical integration of the master equation (solid line)
are in excellent agreement with the results of computer simulations (circles).
They are very well converged towards the
corresponding Poisson distribution $\pi(k|\langle K \rangle_t)$
with the same mean degree $\langle K \rangle_t$ (dashed line).

\subsection{Configuration model networks with exponential degree distributions}

Consider a configuration model network with an exponential degree
distribution of the form $P_0(k) \sim e^{- \alpha k}$, 
where $k \ge k_{\rm min}$
and $k_{\rm min}$ is the lower cutoff of the initial degree distribution.
It is convenient to parametrize the degree distribution 
using the mean degree $\langle K \rangle_0$,
in the form

\begin{equation}
P_0(k) = 
\left\{
\begin{array}{ll}
0    &  \ \ \ \ \   k < k_{\rm min} \\
D \left( \frac{\langle K \rangle_0 - k_{\rm min}}{\langle K \rangle_0 - k_{\rm min} + 1} \right)^k
& \ \ \ \ \   k \ge k_{\rm min},
\end{array}
\right.
\label{eq:exp}
\end{equation}

\noindent
where $D$ is the normalization constant, given by

\begin{equation}
D = 
\frac{1}{(\langle K \rangle_0 - k_{\rm min})+1}
\left( \frac{ \langle K \rangle_0 - k_{\rm min}  }
{ \langle K \rangle_0 - k_{\rm min} + 1 } \right)^{- k_{\rm min}}.
\end{equation}

\noindent
Below we evaluate the relative entropy of an initial network 
with an exponential degree distribution with respect to
the corresponding Poisson distribution.
Inserting the exponential degree distribution of Eq. (\ref{eq:exp})
into Eq. (\ref{eq:Shannon})
and carrying out the summation, we obtain the Shannon entropy

\begin{eqnarray}
S[P_0(k)]
&=& -
\sum_{k=k_{\rm min}}^{\infty} P_0(k) \ln [ P_0(k) ] 
\nonumber \\
&=&
-  (\langle K \rangle_0 - k_{\rm min}) \ln (\langle K \rangle_0 - k_{\rm min}) 
\nonumber \\
&+& (\langle K \rangle_0 - k_{\rm min}+1) \ln (\langle K \rangle_0 - k_{\rm min}+1).
\label{eq:S0exp}
\end{eqnarray}

\noindent
In order to calculate the cross-entropy 
$C[P_0(k) || \pi(k|\langle K \rangle_0)]$, 
we insert the exponential
distribution $P_0(k)$ 
of Eq. (\ref{eq:exp})
into Eq. (\ref{eq:Skk}).
We obtain

\begin{eqnarray}
C[P_0(k) || \pi(k|\langle K \rangle_0)] &=& 
- \langle K \rangle_0 \ln(\langle K \rangle_0)
+ \sum_{k=k_{\rm min}}^{\infty} \left( k + \frac{1}{2} \right) \ln (k)  
\left[  D \left( \frac{\langle K \rangle_0-k_{\rm min}}{\langle K \rangle_0 - k_{\rm min}+1} \right)^k  \right]
\nonumber \\
&+& \frac{1}{2} \ln (2 \pi)
-\frac{1}{2} \ln (2 \pi) 
P_0(0)
+ \left[ 1 - \frac{1}{2} \ln (2 \pi) \right]
P_0(1)
\nonumber \\
&+& \left[ 2 - \frac{3}{2} \ln (2) - \frac{1}{2} \ln (2 \pi) \right] 
P_0(2).
\label{eq:Skke7}
\end{eqnarray}

\noindent
Carrying out the summation, we obtain

\begin{eqnarray}
C[P_0(k) || \pi(k|\langle K \rangle_0)] &=& 
-  \langle K \rangle_0  \ln (\langle K \rangle_0) 
\nonumber \\
&-& \frac{1}{2(\langle K \rangle_0 - k_{\rm min} + 1)}
\left[ 2 \frac{\partial}{\partial \gamma} 
 \Phi
\left. \left( \frac{ \langle K \rangle_0-k_{\rm min} }
{ \langle K \rangle_0-k_{\rm min}+1},\gamma,k_{\rm min} \right) \right|_{\gamma=-1}
\right.
\nonumber \\
&+&
\left.
\frac{\partial}{\partial \gamma} 
\Phi
\left. \left( \frac{ \langle K \rangle_0-k_{\rm min} }
{ \langle K \rangle_0-k_{\rm min}+1},\gamma,k_{\rm min} \right) \right|_{\gamma=0} \right]
\nonumber \\
&+& \frac{1}{2} \ln (2 \pi)
-\frac{1}{2} \ln (2 \pi) 
P_0(0)
+ \left[ 1 - \frac{1}{2} \ln (2 \pi) \right] 
P_0(1)
\nonumber \\
&+& \left[ 2 - \frac{3}{2} \ln (2) - \frac{1}{2} \ln (2 \pi) \right] 
P_0(2),
\label{eq:Skke8}
\end{eqnarray}

\noindent
where $\Phi(x,\gamma,k)$ is the Lerch transcendent
\cite{Olver2010}.
The relative entropy takes the form
$S_0 = -S[P_0(k)] + C[P_0(k)||\pi(k|\langle K \rangle_0)]$,
where $S[P_0(k)]$ is given by Eq. (\ref{eq:S0exp})
and $C[P_0(k)||\pi(k|\langle K \rangle_0)]$
is given by Eq. (\ref{eq:Skke8}).

Below we analyze the convergence of a configuration model network
with an exponential degree distribution towards an ER graph structure upon 
contraction. 
In particular, we calculate the time dependent degree distribution
$P_t(k)$ during contraction and examine its convergence towards $\pi(k|\langle K \rangle_t)$.
To this end we perform direct numerical integration of the 
master equation (\ref{eq:dP/dt}) and computer simulations,
starting from a configuration model network
with a degree distribution
given by Eq. (\ref{eq:exp})
and evaluate the time-dependent relative entropy $S_t$.

\begin{figure}
\begin{center}
\includegraphics[width=5.8cm]{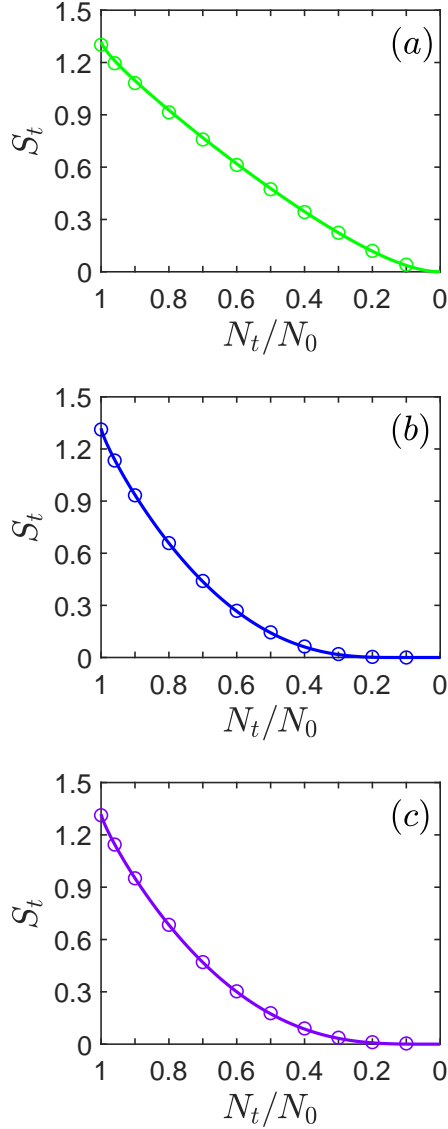}
\caption{
(Color online)
The relative entropy $S_t$ as a function of time 
for a configuration model network 
of initial size $N_0=10^4$ and mean degree $\langle K \rangle_0=20$ 
with an exponential degree distribution in which $k_{\rm min}=10$,
that contracts via random deletion (a), preferential deletion (b) and propagating
deletion (c), obtained from numerical integration of the master equation (solid lines).
In all three cases the relative entropy quickly decays,
which implies that the degree distribution of the contracting network
converges towards a Poisson distribution.
The convergence  is dramatically faster in the preferential and the propagating deletion
scenarios compared to random deletion scenario.
The master equation results 
are in very good agreement with the results obtained from computer simulations (circles).
Also, the initial value $S_0 \simeq 1.32$ is in perfect agreement with 
the result obtained from Eqs. (\ref{eq:S0exp}) and (\ref{eq:Skke8}).
}
\label{fig:4}
\end{center}
\end{figure}

In Fig. \ref{fig:4} we present the
relative entropy $S_t$ as a function of time 
for a configuration model network 
of initial size $N_0=10^4$ and initial mean degree $\langle K \rangle_0=20$
with an exponential degree distribution
that contracts via random deletion (a), preferential deletion (b) and propagating
deletion (c), obtained from numerical integration of the master equation (solid lines).
In all three cases the relative entropy quickly decays,
which implies that the degree distribution of the contracting network
converges towards a Poisson distribution.
The convergence  is dramatically faster in the preferential and the propagating deletion
scenarios compared to random deletion scenario.
The master equation results 
are in very good agreement with the results obtained from computer simulations (circles).

\begin{figure}
\begin{center}
\includegraphics[width=13.4cm]{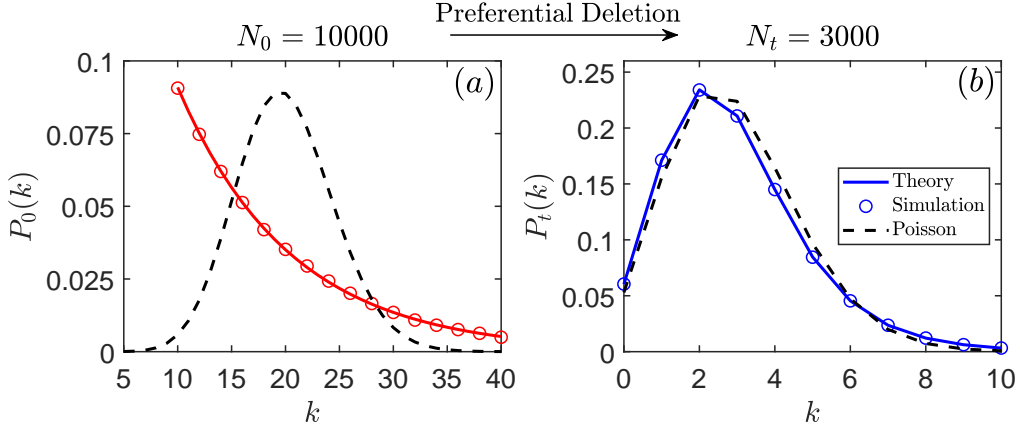}
\caption{
(Color online)
(a) The degree distribution $P_0(k)$ of a configuration model network  
with mean degree $\langle K \rangle_0=20$ 
and an exponential degree distribution,
given by Eq. (\ref{eq:exp}) with $k_{\rm min}=10$ (solid line). 
The circles represent the degree sequence of the $N_0=10^4$ nodes 
in a single realization of the initial network, which was used in the computer simulation.
The corresponding Poisson distribution
with the same mean degree is also shown (dashed line).
The network contracts via the preferential node deletion scenario.
(b) The degree distribution $P_t(k)$ of the contracted network
at time $t=7000$, when the network size is reduced to $N_t=3000$,
obtained from numerical integration of the master equation (solid line).
The master equation results 
are in excellent agreement with the results
obtained from computer simulations (circles).
The corresponding Poisson distribution $\pi(k|\langle K \rangle_t)$
with the same mean degree is also shown (dashed line).
The master equation results and the computer simulation results  
are in very good agreement with the
corresponding Poisson distribution with the same mean degree.
}
\label{fig:5}
\end{center}
\end{figure}

In Fig. \ref{fig:5}(a) we present the degree distribution $P_0(k)$ of a configuration model network  
of size $N_0=10^4$ and an exponential degree distribution with mean degree $\langle K \rangle_0=20$ (solid line).  
The corresponding Poisson distribution
with the same mean degree is also shown (dashed line).
The network contracts via preferential node deletion.
In Fig. \ref{fig:5}(b) we present the degree distribution $P_t(k)$ of the contracted network
at time $t=7000$, when the network size is reduced to $N_t=3000$,
obtained from numerical integration of the master equation (solid line)
and from computer simulations (circles).
The corresponding Poisson distribution $\pi(k|\langle K \rangle_t)$
with the same mean degree is also shown (dashed line).
The master equation results, the computer simulation results and the
corresponding Poisson distribution are found to be in very good agreement
with each other.

\begin{figure}
\begin{center}
\includegraphics[width=6.4cm]{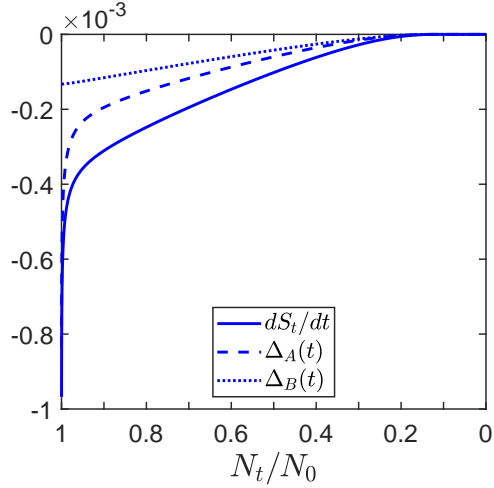}
\caption{
(Color online)
The time derivative of the relative entropy, $dS_t/dt=\Delta_{\rm A}(t)+\Delta_{\rm B}(t)$, 
as a function of time, 
for a configuration model network
of initial size $N_0=10^4$ and exponential degree distribution with mean degree $\langle K \rangle_0=20$
and $k_{\rm min}=10$,
that contracts via preferential node deletion, obtained from numerical integration of the master equation (solid lines).
The terms $\Delta_{\rm A}(t)$ (dashed line) and $\Delta_{\rm B}(t)$ (dotted line),
which sum up to the derivative $dS_t/dt$ are also shown.
Note that both $\Delta_{\rm A}(t)$ and $\Delta_{\rm B}(t)$
are negative at all times during the contraction process.
}
\label{fig:6}
\end{center}
\end{figure}

In Fig. \ref{fig:6} we present the
time derivative of the relative entropy, $dS_t/dt=\Delta_{\rm A}(t)+\Delta_{\rm B}(t)$, 
as a function of time, 
for a configuration model network
of initial size $N_0=10^4$ and exponential degree distribution with mean degree $\langle K \rangle_0=20$
that contracts via preferential node deletion, 
obtained from numerical integration of the master equation (solid lines).
The terms $\Delta_{\rm A}(t)$ (dashed line) and $\Delta_{\rm B}(t)$ (dotted line),
which sum up to the derivative $dS_t/dt$ are also shown.
As expected, both $\Delta_{\rm A}(t)$ and $\Delta_{\rm B}(t)$
are negative at all times during the contraction process.

\subsection{Configuration model networks with power-law degree distributions}

Consider a configuration model network with a power-law degree distribution
of the form 
$P_0(k) \sim k^{-\gamma}$,
where 
$1 \le k_{\rm min} \le k \le k_{\rm max}$.
Here we focus on the case of $\gamma > 2$, in which the
mean degree, 
$\langle K \rangle_0$, 
is bounded even for 
$k_{\rm max} \rightarrow \infty$.
Power-law distributions do not exhibit a typical scale, and are therefore
referred to as scale-free networks.
The normalized degree distribution is given by

\begin{equation}
P_0(k) = 
\left\{
\begin{array}{ll}
0    &   \ \ \ \ \   k<k_{\rm min}    \\
 D \ {k^{-\gamma}}    & \ \ \ \ \  k_{\rm min} \le k \le k_{\rm max} \\
0 & \ \ \ \ \  k > k_{\rm max},
\end{array}
\right.
\label{eq:PLnorm}
\end{equation}

\noindent
where $D$ is the normalization constant, given by

\begin{equation}
D = D(\gamma,k_{\rm min},k_{\rm max}) = \frac{ 1 }{ \zeta(\gamma,k_{\rm min}) - \zeta(\gamma,k_{\rm max}+1) },
\label{eq:PLnormA}
\end{equation}

\noindent
and $\zeta(\gamma,k)$ is the Hurwitz zeta function 
\cite{Olver2010}.
For $2 < \gamma \le 3$ the mean degree is bounded while the second moment,
$\langle K^2 \rangle$, diverges in the limit of
$k_{\rm max} \rightarrow \infty$.
For $\gamma > 3$ both moments are bounded.
The mean degree is given by

\begin{equation}
\langle K \rangle_0 = 
\frac{ \zeta(\gamma-1,k_{\rm min}) - \zeta(\gamma-1,k_{\rm max}+1) }
{ \zeta(\gamma,k_{\rm min}) - \zeta(\gamma,k_{\rm max}+1) }.
\label{eq:Kmsf}
\end{equation}

\noindent
The second moment of the degree distribution, when finite, is

\begin{equation}
\langle K^2 \rangle_0 = 
\frac{ \zeta(\gamma-2,k_{\rm min}) - \zeta(\gamma-2,k_{\rm max}+1) }
{ \zeta(\gamma,k_{\rm min}) - \zeta(\gamma,k_{\rm max}+1) }.
\label{eq:K2msf}
\end{equation}

Below we evaluate the relative entropy of an initial network 
with a power law degree distribution with respect to
the corresponding Poisson distribution.
In order to calculate the Shannon entropy
$S[P_0(k)]$ we insert the power-law distribution of Eq. (\ref{eq:PLnorm})
into Eq. (\ref{eq:Shannon}).
We obtain

\begin{equation}
S[P_0(k)] = 
-
\sum_{k=k_{\rm min}}^{\infty} 
P_0(k) \ln [ P_0(k) ] = 
-
\ln (D) + \gamma \sum_{k=k_{\rm min}}^{\infty}
D k^{- \gamma} \ln (k).
\label{eq:PLPsf}
\end{equation}

\noindent
Since $\ln (1) = 0$ the summation in Eq. (\ref{eq:PLPsf})
actually starts from the larger value between $k=2$ and $k_{\rm min}$,
denoted by $\overline k_{\rm min}=\max \{2,k_{\rm min} \}$.
We thus obtain

\begin{equation}
S[P_0(k)] = 
-\ln (D) + 
\gamma \sum_{ k= \overline k_{\rm min} }^{\infty}
D k^{- \gamma} \ln (k).
\label{eq:PLPsfi}
\end{equation}
 
\noindent
Carrying out the summation, we obtain

\begin{equation}
S[P_0(k)] = 
- \ln (D)  
+ \gamma D \left[  \zeta'(\gamma,k_{\rm max}+1)   -   \zeta'(\gamma,\overline k_{\rm min}) \right],
\label{eq:PLPsfis}
\end{equation}

\noindent
where 
$\zeta'(\gamma,k) = \partial \zeta(\gamma,k)/\partial \gamma$.

In order to calculate the cross-entropy 
$C[P_0(k) || \pi(k|\langle K \rangle_0)]$, 
we insert the power-law
distribution $P_0(k)$ into Eq. (\ref{eq:Skk}).
We obtain

\begin{eqnarray}
C[P_0(k) || \pi(k|\langle K \rangle_0)] &=& 
- \langle K \rangle_0  \ln (\langle K \rangle_0) 
+ \sum_{k = \overline k_{\rm min}}^{\infty} \left( k + \frac{1}{2} \right) \ln (k)  D k^{-\gamma}
\nonumber \\
&+& \frac{1}{2} \ln (2 \pi)
+ \left[ 1 - \frac{1}{2} \ln (2 \pi) \right] P_0(1)
\nonumber \\
&+& \left[ 2 - \frac{3}{2} \ln (2) - \frac{1}{2} \ln (2 \pi) \right] P_0(2).
\label{eq:Skk17}
\end{eqnarray}

\noindent
Carrying out the summation, we obtain

\begin{eqnarray}
C[P_0(k) || \pi(k|\langle K \rangle_0)] &=& 
- \langle K \rangle_0  \ln (\langle K \rangle_0) 
+ 
D \left[ \zeta'(\gamma-1,\overline k_{\rm min}) - \zeta'(\gamma-1,k_{\rm max}+1) \right]
\nonumber \\
&+& \frac{D}{2} 
\left[ \zeta'(\gamma, k_{\rm min}) - \zeta'(\gamma,k_{\rm max}+1) \right]
\nonumber \\
&+& \frac{1}{2} \ln (2 \pi)
+ \left[ 1 - \frac{1}{2} \ln (2 \pi) \right] 
P_0(1)
\nonumber \\
&+& \left[ 2 - \frac{3}{2} \ln (2) - \frac{1}{2} \ln (2 \pi) \right] 
P_0(2).
\label{eq:Skk8}
\end{eqnarray}

\noindent
The relative entropy 
of the initial network with a power-law degree distribution given by Eq. (\ref{eq:PLnorm})
takes the form
$S_0 = -S[P_0(k)] + C[P_0(k)||\pi(k|\langle K \rangle_0)]$,
where $S[P_0(k)]$ is given by Eq. (\ref{eq:PLPsfis})
and $C[P_0(k)||\pi(k|\langle K \rangle_0)]$
is given by Eq. (\ref{eq:Skk8}).

Below we analyze the convergence of a configuration model network
with a power-law degree distribution towards an ER graph structure upon 
contraction. 
In particular, we calculate the time dependent degree distribution
$P_t(k)$ during contraction and examine its convergence towards $\pi(k|\langle K \rangle_t)$.
To this end we perform direct numerical integration of the 
master equation (\ref{eq:dP/dt}) and computer simulations,
starting from a configuration model network
with a degree distribution
given by Eq. (\ref{eq:PLnorm})
and evaluate the time-dependent relative entropy $S_t$.

\begin{figure}
\begin{center}
\includegraphics[width=5.8cm]{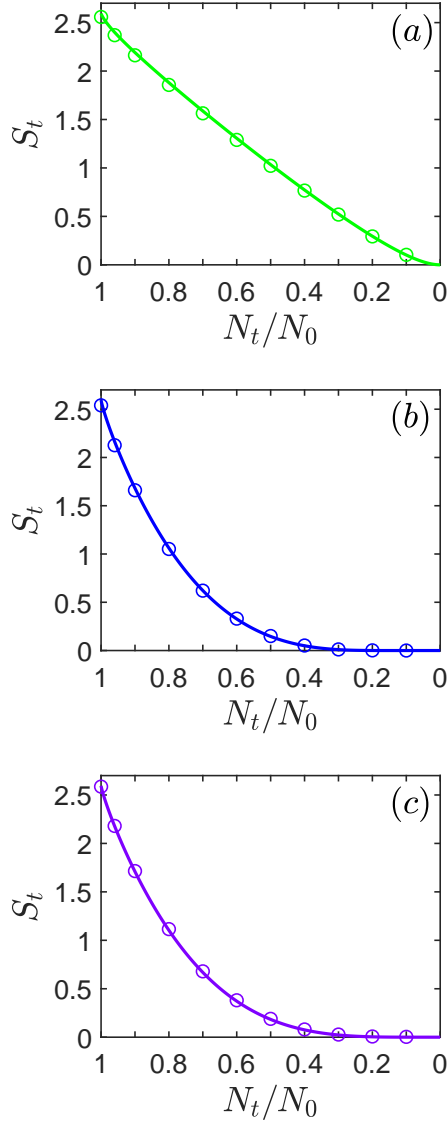}
\caption{
(Color online)
The relative entropy $S_t$ as a function of time 
for a configuration model network 
with a power-law degree distribution
of initial size $N_0=10^4$ and mean degree $\langle K \rangle_0=20$,
where $k_{\rm min}=10$, $k_{\rm max}=100$ and $\gamma=2.65$,
that contracts via random deletion (a), preferential deletion (b) and propagating
deletion (c), obtained from numerical integration of the master equation (solid lines).
In all three cases the relative entropy quickly decays,
which implies that the degree distribution of the contracting network
converges towards a Poisson distribution.
The convergence  is dramatically faster in the preferential and the propagating deletion
scenarios compared to random deletion scenario.
The master equation results 
are in very good agreement with the results obtained from computer simulations (circles).
Also, the initial value $S_0 \simeq 2.59$ is in perfect agreement with 
the result obtained from Eqs. (\ref{eq:PLPsfis}) and (\ref{eq:Skk8}).
}
\label{fig:7}
\end{center}
\end{figure}
  
In Fig. \ref{fig:7} we present the 
relative entropy $S_t$ as a function of time 
for a configuration model network 
with a power-law degree distribution,
of initial size $N_0=10^4$ and initial mean degree $\langle K \rangle_0=20$,
where $k_{\rm min}=10$, $k_{\rm max}=100$ and $\gamma=2.65$,
that contracts via random deletion (a), preferential deletion (b) and propagating
deletion (c), obtained from numerical integration of the master equation (solid lines).
In all three cases the relative entropy quickly decays,
which implies that the degree distribution of the contracting network
converges towards a Poisson distribution.
The convergence  is dramatically faster in the preferential and the propagating deletion
scenarios compared to random deletion scenario.
The master equation results 
are in very good agreement with the results obtained from computer simulations (circles).

\begin{figure}
\begin{center}
\includegraphics[width=13.4cm]{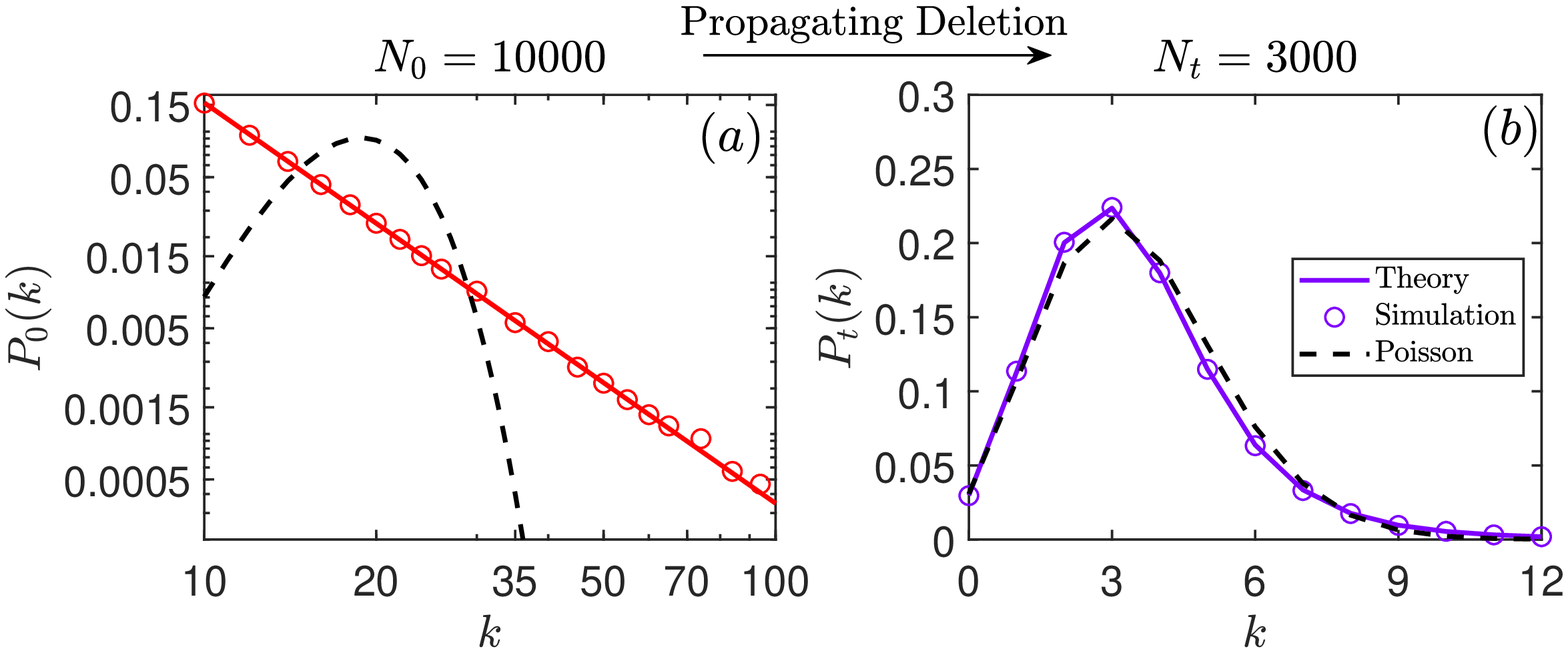}
\caption{
(Color online)
(a) The degree distribution $P_0(k)$ of a configuration model network  
with a power-law degree distribution,
given by Eq. (\ref{eq:PLnorm}), and mean degree 
$\langle K \rangle_0=20$ (solid line),
where $k_{\rm min}=10$, $k_{\rm max}=100$ and $\gamma=2.65$,
is shown on a log-log scale. 
The circles represent the degree sequence of the $N_0=10^4$ nodes 
in a single realization of the initial network, which was used in the computer simulation.
The corresponding Poisson distribution
with the same mean degree is also shown (dashed line).
The network contracts via the propagating node deletion scenario.
(b) The degree distribution $P_t(k)$ of the contracted network
at time $t=7000$, when the network size is reduced to $N_t=3000$,
obtained from numerical integration of the master equation
is shown on a linear scale (solid line).
The master equation results 
are in excellent agreement with the results
obtained from computer simulations (circles).
The corresponding Poisson distribution $\pi(k|\langle K \rangle_t)$
with the same mean degree is also shown (dashed line).
The master equation results and the computer simulation results  
are in very good agreement with the
corresponding Poisson distribution with the same mean degree.
}
\label{fig:8}
\end{center}
\end{figure}

In Fig. \ref{fig:8}(a) we present the degree distribution $P_0(k)$ of a configuration model network  
of size $N_0=10^4$ and a power-law degree distribution with mean degree $\langle K \rangle_0=20$ (solid line).  
The corresponding Poisson distribution
with the same mean degree is also shown (dashed line).
The network contracts via propagating node deletion.
In Fig. \ref{fig:8}(b) we present the degree distribution $P_t(k)$ of the contracted network
at $t=7000$, when the network size is reduced to $N_t=3000$,
obtained from numerical integration of the master equation (solid line)
and from computer simulations (circles).
The corresponding Poisson distribution $\pi(k|\langle K \rangle_t)$
with the same mean degree is also shown (dashed line).
The master equation results, the computer simulation results and the
corresponding Poisson distribution are found to be in very good agreement
with each other.

\begin{figure}
\begin{center}
\includegraphics[width=6.4cm]{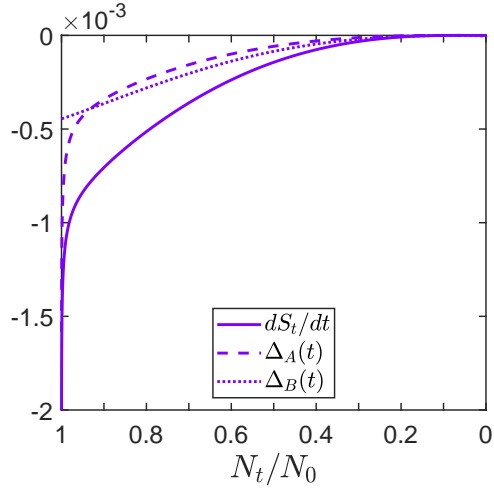}
\caption{
(Color online)
The time derivative of the relative entropy $dS_t/dt$
as a function of time, 
for a configuration model network
of initial size $N_0=10^4$ and a power-law degree distribution with mean degree $\langle K \rangle_0=20$,
where $k_{\rm min}=10$, $k_{\rm max}=100$ and $\gamma=2.65$,
that contracts via propagating node
deletion, obtained from numerical integration of the master equation (solid lines).
The terms $\Delta_{\rm A}(t)$ (dashed line) and $\Delta_{\rm B}(t)$ (dotted line),
which sum up to the derivative $dS_t/dt$ are also shown.
Note that both $\Delta_{\rm A}(t)$ and $\Delta_{\rm B}(t)$
are negative at all times during the contraction process.
}
\label{fig:9}
\end{center}
\end{figure}

In Fig. \ref{fig:9} we present the 
time derivative of the relative entropy, $dS_t/dt$
as a function of time, 
for a configuration model network
of initial size $N=10^4$ and a power-law degree distribution with mean degree $\langle K \rangle_0=20$
that contracts via propagating node
deletion, obtained from numerical integration of the master equation (solid lines).
The terms $\Delta_{\rm A}(t)$ (dashed line) and $\Delta_{\rm B}(t)$ (dotted line),
which sum up to the derivative $dS_t/dt$ are also shown.
As expected, both $\Delta_{\rm A}(t)$ and $\Delta_{\rm B}(t)$
are negative at all times during the contraction process.

\section{Discussion}

In Ref. 
\cite{Tishby2019}
we used direct numerical integration of the master equation and computer simulations 
to show that the degree distributions of 
contracting networks converge towards the Poisson distribution. 
To this end, we used the relative entropy as a distance measure between 
the degree distribution $P_t(k)$ of the contracing network and the
corresponding Poisson distribution $\pi(k|\langle K \rangle_t)$, and showed that this
distance decreases as the network contracts.

A computer simulation of network contraction provides results for
a single instance of the initial network and a single stochastic path of 
the contraction process. In order to obtain statistically significant results 
for a given ensemble of initial networks and given network contraction scenario
one needs to combine the results of a large number of independent runs.
The direct numerical integration of the master equation is advantageous in the sense
that a single run of the numerical integration process provides results
for a whole ensemble of initial networks.
However, a given network ensemble represents a single point in the 
high dimensional parameter space of possible network ensembles.
Therefore, in order to explore the general properties of network contraction
processes one needs to repeatedly apply the direct integration of the master equation
to a large sample of distinct network ensembles.

Our aim in this paper was to obtain rigorous analytical results for the
convergence of contracting networks towards the ER
network ensemble.  To this end we devised a rigorous argument, which is based
on the master equation that describes the temporal evolution of the
degree distribution $P_t(k)$ and the relative entropy $S_t$.
Such an argument is advantageous over the direct numerical integration of
the master equation or
computer simulations in the sense that it is universally applicable to all
possible degree distributions.
%To analyze the time dependence of the relative entropy $S_t$, we 
%expressed its time derivative as a sum of two terms, $\Delta_{\rm A}(t)$
%and $\Delta_{\rm B}(t)$.
%The $\Delta_{\rm A}(t)$ term is associated with the 
%trickle-down term of the master equation, which accounts for 
%secondary effect of node deletion on the degrees of the neighbors of the deleted node,
%while the $\Delta_{\rm B}(t)$ term is associated with the redistribution term,
%which accounts for the direct effect of the removal of the deleted node.
%We have shown that the first term satisfies $\Delta_{\rm A}(t) < 0$
%for any degree distribution $P_t(k)$.
%This means that the $\Delta_{\rm A}(t)$ term always pushes the relative entropy down towards zero,
%driving the convergence of $P_t(k)$ towards Poisson.
%For the $\Delta_{\rm B}(t)$ term we provide a condition
%that can be used for any given degree distribution $P_t(k)$ to determine
%whether this term would accelerate the convergence to Poisson or slow it down.
%The condition implies that for degree distributions $P_t(k)$ whose tail falls more slowly than
%the tail of the corresponding Poisson distribution, the $\Delta_{\rm B}(t)$ term
%would accelerate the convergence to Poisson, while in case that the tail falls more
%quickly than Poisson the $\Delta_{\rm B}(t)$ term whould slow down the convergence.

The relative entropy $S[P(k) || Q(k)]$ of a distribution $P(k)$ with 
respect to a distribution $Q(k)$ is a special case of the R\'enyi divergence
$S_{\alpha}[P(k) || Q(k)]$, with $\alpha=1$
\cite{Renyi1961}.
The choice of $\alpha=1$ is advantageous in the sense that it has
a natural information theoretic interpretation
\cite{Annibale2009,Roberts2011}.
The relative entropy is an asymmetric
distance measure, or quasi-distance
\cite{Deza2016}.
Interestingly, the relative entropy
is related to other distance measures between discrete probability distributions.
For example, the total variation distance between probability distributions
$P(k)$ and $Q(k)$ is given by
$T[ P(k), Q(k)] = \sum_{k} | P(k) - Q(k) |$,
namely, the sum of the differences (in absolute value) between the 
probabilities assigned to all values of $k$ by the two distributions.
Clearly, for any two distributions $P(k)$ and $Q(k)$, the total variation distance
satisfies
$0 \le T[ P(k), Q(k)] \le 2$.
The relative entropy provides an additional upper bound on the total variation
distance via the Pinsker inequality, which takes the form
\cite{Pinsker1964,Kullback1966,Csiszar1967,Vajda1970}

\begin{equation}
T[ P(k), Q(k)]  \le \sqrt{ \frac{1}{2} S[P(k) || Q(k)] }.
\end{equation}

\noindent
This relation implies that whenever the relative entropy between
$P(k)$ and $Q(k)$ vanishes, so does the total variation distance 
between them, meaning that the two distributions become identical 
in the $L_1$ norm. This shows that when the relative entropy vanishes 
the distributions become identical.

In this paper we focused on the case of configuration model networks, which exhibit
a given degree distribution and no degree-degree correlations. The
theoretical framework presented here may provide the foundations for 
the study of network contraction processes in a much broader class of 
complex networks, which exhibit degree-degree correlations as well 
as other structural correlations. 
This will require a more general formulation of the relative entropy, expressed
in terms of the joint degree distributions of pairs or adjacent nodes,
which take into account the correlations between their degrees.

The theoretical framework presented here may be relevant in the broad
context of neurodegeneration, which is the progressive loss
of structure and function of neurons in the brain.
Such processes occur in 
normal aging 
\cite{Morrison1997}
as well as in
a large number of incurable neurodegenerative
diseases such as Alzheimer, Parkinson, Huntington and Amylotrophic
Lateral Sclerosis, which result in a gradual loss of cognitive and
motoric functions
\cite{Heemels2016}. 
These diseases differ in the specific brain regions or circuits
in which the degeneration occurs.
The characterization of the evolving structure 
using the relative entropy
may provide useful insight into 
the structural aspects of
the loss of neurons and synapses in
neurodegenerative processes
\cite{Arendt2015}.

It is worth mentioning that there is
another class of network dismantling processes that involve 
optimized attacks, which maximize the damage to the network for a 
minimal set of deleted nodes
\cite{Braunstein2016,Zdeborova2016}.
Such optimization is achieved by first decycling the network, 
namely, by selectively deleting nodes that reside on cycles, thus
driving the giant component into a tree structure. 
The branches of the tree are then trimmed such
that the giant component is quickly disintegrates.
Clearly, these optimized dismantling processes do not
converge towards an ER structure.

\section{Summary}

In summary, we have analyzed the structural evolution of 
complex networks undergoing contraction processes via 
generic node deletion scenarios, namely,
random deletion, preferential deletion
and propagating deletion.
Focusing on configuration model networks 
we have shown using a rigorous argument
that upon contraction the degree distributions of these
networks converge towards a
Poisson distribution.
In this analysis we used
the relative entropy $S_t=S[P_t(k) || \pi(k|\langle K \rangle_t)]$ 
of the degree distribution $P_t(k)$ of the contracting
network at time $t$ with respect to the corresponding Poisson distribution
$\pi(k|\langle K \rangle_t)$ with the same mean degree $\langle K \rangle_t$
as a distance measure between $P_t(k)$ and Poisson.
The relative entropy 
is suitable as a distance measure since it
satisfies $S_t \ge 0$ for any degree
distribution $P_t(k)$, while equality is obtained only for 
$P_t(k) = \pi(k|\langle K \rangle_t)$.
We derived an equation for the time evolution
of the relative entropy $S_t$
during network contraction 
and expressed its time derivative $dS_t/dt$ as a sum of two terms, $\Delta_{\rm A}(t)$
and $\Delta_{\rm B}(t)$.
We have shown that the first term satisfies $\Delta_{\rm A}(t) < 0$
for any degree distribution $P_t(k)$.
This means that the $\Delta_{\rm A}(t)$ term always pushes the relative entropy down towards zero,
driving the convergence of $P_t(k)$ towards Poisson.
For the $\Delta_{\rm B}(t)$ term we provide a condition
that can be used for any given degree distribution $P_t(k)$ to determine
whether this term would accelerate the convergence to Poisson or slow it down.
The condition implies that for degree distributions $P_t(k)$ whose tail falls more slowly than
the tail of the corresponding Poisson distribution, the $\Delta_{\rm B}(t)$ term
would accelerate the convergence to Poisson, while in the case that the tail falls more
quickly than Poisson the $\Delta_{\rm B}(t)$ term whould slow down the convergence.
We analyzed the convergence for configuration model networks with 
degenerate degree distributions (random regular graphs), exponential
degree distributions and power-law degree distributions (scale-free networks) 
and showed that the relative entropy 
decreases monotonically to zero during the contraction process,
reflecting the 
convergence of the degree distribution towards a Poisson distribution.
Since the contracting networks remain uncorrelated,
this means that their structures converge towards an
Erd{\H o}s-R\'enyi (ER) graph structure,
substantiating earlier results obtained using
direct integration of the master equation and computer simulations
\cite{Tishby2019}.

This work was supported by the Israel Science Foundation grant no. 
1682/18.

\end{document}